\documentclass[a4paper,11pt]{article}

\pdfoutput=1

\usepackage{jheppub}
\usepackage{amsfonts}
\usepackage{graphicx,color}
\usepackage{float}
\usepackage{hyperref}
\usepackage{subfigure}
\usepackage[utf8]{inputenc}
\usepackage[english]{babel}
\usepackage{textcomp}
\usepackage{mathtools}
\usepackage{gensymb}
\usepackage{amsbsy}
\usepackage{capt-of}
\usepackage[T1]{fontenc} 
\usepackage{physics}
\usepackage{multicol} 
\usepackage{multirow}
\usepackage[table]{xcolor}
\usepackage{lscape}
\usepackage[format=plain,small,labelfont={bf},up]{caption} 
\usepackage[title]{appendix}
\usepackage{graphicx}  
\usepackage{subcaption} 
\usepackage{amsmath,amsthm,amssymb}
\usepackage{soul}
\usepackage{blkarray}
\usepackage{array}
\usepackage{booktabs} 

\RequirePackage{color}
\usepackage{colortbl}
\definecolor{denim}{rgb}{0.08, 0.38, 0.74}
\hypersetup{
    colorlinks=true,
    linkbordercolor={white},
    linkcolor={denim},
    citecolor={denim},
    filecolor={denim},      
    urlcolor={denim},
}

\numberwithin{equation}{section}

\def\be{\begin{equation}}
\def\ee{\end{equation}}
\def\bea{\begin{eqnarray}}
\def\eea{\end{eqnarray}}

\def\eg{{\it e.g.}}

\def\ie{{\it i.e.}}

\setcitestyle{square, numbers, comma, sort&compress}

\title{\boldmath Warm Warped Throats}

\author[a]{Dibya Chakraborty}    
\author[b]{and Rudnei O. Ramos}

\affiliation[a]{Centre for Strings, Gravitation and Cosmology, Department of Physics, Indian Institute of
Technology Madras, Chennai 600036, India}
\affiliation[b]{Departamento de F\'{\i}sica Te\'orica,
  Universidade do Estado do Rio de Janeiro,
  20550-013 Rio de Janeiro, RJ, Brazil}

\emailAdd{dibyac@fisica.ugto.mx}
\emailAdd{rudnei@uerj.br}

\abstract{
We investigate brane inflation, focusing on warm inflation realizations within 
a warped throat geometry. While the standard scenario relies on a single mobile 
$D3$-brane moving radially toward an anti-$D3$-brane at the throat's tip, we 
propose two distinct single-field inflationary pictures. In our approach, the radial and angular 
coordinates of a $D3$-brane on a warped deformed conifold act as two independent 
inflaton fields. We address moduli stabilization by incorporating a supersymmetrically 
embedded $D7$-brane, which generates the necessary radial and angular scalar potentials. 
Evaluating these radial and angular brane inflation setups within the warm inflation 
paradigm, we demonstrate that dissipation effects allow the models to satisfy recent 
observational constraints more naturally than their cold inflation counterparts for 
a given parameter space.
}


\arxivnumber{2603.29781}

\begin{document} 
 \maketitle

\section{Introduction}

The standard recipe for an inflationary model in string theory is to fix all heavy moduli such as complex structures and dilaton using supersymmetry (SUSY) conditions of type-IIB supergravity~\cite{Kachru:2003aw}. This leaves K\"ahler moduli as flat directions. Next, one adds non-perturbative (NP) effects such as Euclidean $D3$-branes or gaugino condensation on a stack of 
$D7$-branes wrapping a divisor inside the Calabi-Yau (CY). This generates K\"ahler moduli dependent superpotentials, which in turn produces supersymmetric anti-de Sitter (AdS) minimum. As a final step, an anti-brane effect is added to finally lift it to a non-supersymmetric dS potential. This procedure describes, for example, the standard procedure for the KKLT setup~\cite{Kachru:2003aw}. 
Although this is the standard picture of a stringy inflationary model, the brane inflation studied earlier~\cite{Dvali:1998pa}  follows a different route.

The original proposal for brane inflation was proposed in~\cite{Dvali:1998pa}, where a brane-antibrane was considered and the distance between them played the role of a scalar field driving inflation. The energy driving inflation is sourced via Coulomb interactions between brane-antibrane pairs. Later, the KKLMMT model was proposed~\cite{Kachru:2003sx}, setting an important step forward, as the authors incorporated the effects of the setup of original brane-antibrane inflation \cite{Dvali:1998pa} embedded in type IIB flux compactification~\cite{Giddings:2001yu,Kachru:2003aw}. The importance of a warped throat\footnote{Note that, by warped throat we mean a deformed conifold whose cycles are threaded by fluxes from type-IIB compactifications. Hence, we shall interchangeably use both notations of warped deformed conifold and warped throat to mean the same geometry.} was first emphasized in~\cite{Dvali:1998pa} in which it helps to flatten the brane potential suitable for a prolonged stage of inflation. However, once the effect of moduli stabilization via the F-term potential was incorporated, it generates some non-trivial terms to the mass term of the inflaton, inducing the known $\eta-$problem of inflation. {}Finally, in~\cite{Krause:2007jk} and later in~\cite{Pajer:2008uy}, a simpler and more realistic model of brane inflation was proposed incorporating the effect of moduli stabilization. 

Since moduli stabilization lies at the core of any string-theoretic construction, it is more desirable to build inflationary models that incorporate the effects of compactification. This was achieved in~\cite{Krause:2007jk} for radial brane inflation and in~\cite{Pajer:2008uy} for angular brane inflation. The $\eta-$problem of inflation faced by moduli stabilization effect can be ameliorated by the presence of inflaton dependent threshold corrections to the superpotential. These corrections to the superpotential were made available for the warped conifold background in  ref.~\cite{Baumann:2006th}. Threshold corrections modify the superpotential, making it proportional to $f(z)^{1/n}$, 
where $f(z)$ is the functional taking care of the NP term in the superpotential and with
$z$ denotes the complex coordinates of the $3-$sphere of the deformed conifold at the tip, which is also the coordinates of the $D3$-brane and $n$ denotes the number of coinciding $D7$-branes.

Anti-brane in our work is either absent or sufficiently far away. Hence, the model bypasses the need for additional symmetries or fine tuning. The Coulomb potential is absent in the case of angular brane inflation and is present in the radial brane but is sub-dominant as expected~\cite{Krause:2007jk}. In the usual scenario, it is the brane-anti-brane annihilation that ends inflation. In the absence of anti-brane, some other mechanism must be responsible for stopping inflation and also for producing reheating. As stated in~\cite{Pajer:2008uy} and shown in~\cite{Kofman:2004yc}, brane-brane collisions, in which the kinetic energy of the inflating brane is completely transferred to the newly created stack of branes, can naturally lead to a mechanism for reheating. In our setup, where the tip of the throat contains several $D3$-branes, the brane collision mechanism is easily favorable. The inflation in this setup ends when the field gets trapped in its geometrical minimum by Hubble drag from an inflating Universe. 

The models that we investigate here consist of a space filling $D3$-brane. In case of radial brane inflation, the brane moves along the radial direction of a warped throat. In the case of angular brane inflation, the brane moves along the tip of a warped deformed conifold (WDC), which is a $3-$sphere. Using a symmetric $D7$-brane embedding, one can write the scalar potential as a function of a single scalar corresponding to one angular direction instead of three angular directions of $S^3$ (see \cite{Agarwal:2011wm} for a realization of multi-field brane inflation). We have found that in both  the case of radial and angular brane inflation, in different and broad regions of parameter space, the models can produce viable models of inflation. We do our analysis for both the cases of cold inflation (CI) and warm inflation (WI) and study the appropriate background dynamics. 

Relevant observational quantities, such as the tensor-to-scalar ratio $r$ and the spectral tilt $n_s$ of the power spectrum, are also determined, which help us constrain the different parameters of the models. We emphasize on the fact that the angular and radial brane inflation discussed in this paper are two distinct models. While studying radial brane, one assumes that all the angular coordinates are stabilized at their respective minima, including the conifold modulus, representing the radius of the $S^3-$sphere at the tip of a WDC and do not participate in the dynamics of inflation. The dynamics is entirely driven by the radial distance. The same goes for the angular brane inflationary case, where the other  angular cooridinates, radial distance as well as the conifold modulus are stabilized at their minima. This significantly simplifies the analysis, since then the dynamics reduces to that of a single scalar field coupled to the radiation bath in WI.

WI acts like an important step towards developing an alternative framework of CI. In contrast to CI, as inflation progresses, the inflaton dissipates its energy to a thermal bath, bypassing the need for a separate reheating phase. Another attractive feature of WI is that it can align with an  effective field theory approach and can be consistently embedded into a complete ultraviolet (UV) theory such as string theory~\cite{Das:2018rpg,Motaharfar:2018zyb,Das:2019acf,Berera:2019zdd,Kamali:2019xnt,Berera:2020iyn,Das:2020xmh,Brandenberger:2020oav,Motaharfar:2021egj,Chakraborty:2025yms,Chakraborty:2025jof}. It should also be emphasized that WI has certain distinct signatures that makes it distinguishable from CI, such as non-Gaussianities~\cite{Bastero-Gil:2014raa} and a different consistency relation between the tensor-to-scalar ratio and the tilt of the tensor spectrum~\cite{Bartrum:2013fia}
(for a recent review, see, e.g. ref.~\cite{Kamali:2023lzq}).

The study of brane inflation in the context of WI was first started in \cite{Bastero-Gil:2011zxb},  where the original brane-anti-brane inflation of \cite{Dvali:1998pa} was considered. The authors of ref.~\cite{Bastero-Gil:2011zxb} have found a successful model of WI using the distance between brane-anti-brane as their inflaton candidate. The dissipation mechanism was motivated via the standard two-stage dissipation mechanism introduced in~\cite{Berera:2002sp} (see also ref.~\cite{Berera:2008ar}), according to which the inflaton is coupled to heavy intermediate fields, which in turn are coupled to lighter fields. We have considered the same dissipative mechanism because the radial field still corresponds to a distance in the warped throat. In case of angular brane inflation, since we can compare the inflaton candidate to an axion, we consider the standard axion-like coupling between the inflaton and the standard model (SM) gauge field sector, whose dissipation mechanism resembles that from minimal warm inflation~\cite{Berghaus:2019whh}. 

In this paper, we work with natural units, where we set the speed of light, the Planck and the Boltzmann constants to unit, \ie, $c=\hbar=k_{B}=1$. We use the reduced Planck mass defined as $M_{\rm Pl}=(8\pi G)^{-1/2}\simeq 2.44\times 10^{18}\,\mathrm{GeV},$ where $G$ is the Newton's gravitational constant.

The paper is organized as follows. In section \ref{sec:2}, we briefly review the notable features of a warped throat where the brane moves to give inflation. We briefly discuss  supersymmetric embedding, which plays a pivotal role in the derivation of the potential driving inflation. In section \ref{sec:3}, we discuss the basics of WI and the different dissipation mechanisms that we consider to investigate radial and angular brane inflation in warped throat.  In section \ref{sec:4}, we present our numerical results and contrast them with the observational data. Finally, in section \ref{sec:5}, we conclude and give our final remarks. An appendix is also included to show some of the technical details
about the warped deformed conifold.

\section{$D3$-brane at the tip}
\label{sec:2}

\begin{figure}[H]
    \centering
    \includegraphics[width=0.8\linewidth]{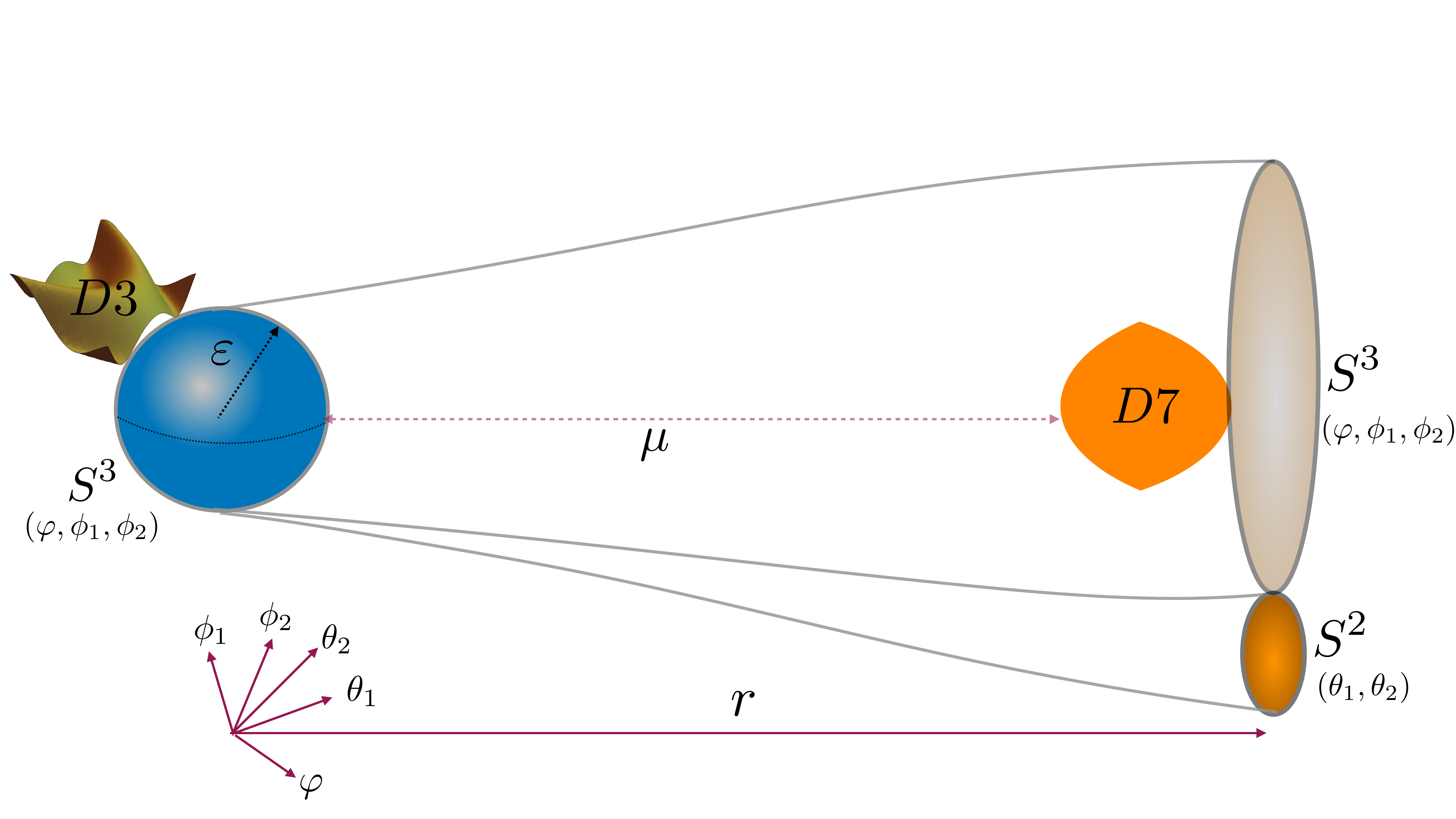}
    \caption{Illustration of the warped throat, for details, see the text. }
    \label{fig:illustration1}
\end{figure}

One fundamental difference in this type of warped brane inflation is the direction along which the $D3$-brane moves. In the standard set-up, it propagates along the radial direction of a warped throat; however, as shown in  \cite{DeWolfe:2007hd,Pajer:2008uy} the authors have encouraged the possibility where the space-filling $D3$-brane moves along one of the angular directions of a $S^3$ sphere at the tip of a throat (see figure \ref{fig:illustration1}). Ideally, the potential depends on three scalars, three angles of a $S^3$ and its potential depends on the F-term contribution coming from supersymmetric embedding of a stack of $D7$-branes. In this section, we plan to look at the anatomy of the radial and angular brane inflation by looking deep into the WDC. Next, by laying out the recipe for the full moduli stabilization, we plan to look at the $D7$-embeddings, as their choice gives rise to the scalar potential of the brane moduli. Let us start by illustrating the motion of the mobile $D3$-brane inside the WDC (see appendix \ref{appA}).

To better understand the motion of the brane inside the conifold, we refer the reader to figure \ref{fig:illustration1}. The cone like structure denotes a conifold, which defines a local structure of a CY manifold. The conifold consists of a base which has $(SU(2)\times SU(2))/U(1)$ symmetry and a radial direction (along $\mu$) corresponds to the height of the throat. The base denotes the $S^3\times S^2$ in the figure \ref{fig:illustration1}. The conifold usually comes with a singularity at the tip \cite{Candelas:1989ug} which can be resolved by blowing up a $S^3$ at the tip as shown above. The new structure without singularity is called a deformed conifold \cite{Candelas:1989js}. Finally, in the presence of type-IIB fluxes to address moduli stabilization, a new name, WDC, is attributed to the geometry. We study our brane inflation in a WDC. The inflaton candidate for the radial brane inflation is the radial coordinate of this conifold, which is along $\mu$. $\mu^{2/3}$ denotes the distance between the tip of the warped throat and the position, where $D7$-brane attaches to the bulk geometry. The radial inflaton is free to move in this distance. On the other hand, in the case of angular inflation, the $D3$-brane is localized on the $S^3$, at the tip of the WDC and moves along a single angular direction, while the remaining four angular directions and the radial direction are stabilized at their respective minima. $\varepsilon$ denotes the radius of the $3-$sphere, also known as conifold modulus. We emphasize that radial and angular brane inflation do not occur simultaneously in the WDC. This allows us to treat them as two distinct inflationary scenarios.

\subsection{Full moduli stabilization}

As stated earlier, one of the main motivations to go beyond the standard brane-inflation is that, in this work, we shall consider the effect of moduli stabilization in type IIB flux compactifications. Following the standard recipe of F-theory, we will in particular focus on the threshold corrections added to the Gukov-Vafa-Witten (GVW) superpotential~\cite{Gukov:1999ya,Gukov:1999gr} and how it depends on the brane moduli. These corrections modifies the standard exponential non-perturbative (NP) effects, which are usually utilized to stabilize the K\"ahler moduli. The moduli stabilization is usually conducted in two steps \cite{Polchinski:1998rr}. As a first step, the complex structure and dilaton are stabilized by fluxes \cite{Lust:1989tj} using SUSY conditions. In the presence of one K\"ahler modulus $\rho$, the $N=1$ supergravity scalar potential takes the form of:
\begin{equation}\label{scalar_pot_gen}
V_F=e^{k_4^2\,K}\left(K^{\overline{I}J}D_JW\overline{D_I W}-3k_4^2|W|^2\right),
\end{equation}
where $K$ is the K\"ahler potential, $W$ denotes the Gukov-Vafa-Witten (GVW) superpotential \cite{Gukov:1999ya,Gukov:1999gr}, and the indices $I,J$ run over the K\"ahler modulus and the open string moduli $z_i$ describing the $D3$-brane position. The K\"ahler potential $K$ in eq.~(\ref{scalar_pot_gen}) is given by \cite{Bagger:1990qh}:
\begin{equation}\label{kahler_pot_gen}
    k_4^2K=-3\,\log\left[\rho+\overline{\rho}-\gamma k(z,\Bar{z})\right]\equiv -3\log U,
\end{equation}
where $\gamma$ is a constant and $k$ denotes the K\"ahler potential of the CY in the position of the $D3$-brane. $U$ is expressed as $U=\rho+\bar{\rho}-\gamma k_0$, with $k_0$ being the value of $k(z,\bar{z})$ at the tip of a WDC as described in the appendix \ref{appA}. $\rho$ is the complex scalar with saxion denoted by $\sigma$ and the axion is denoted by $\theta$, $\rho=\sigma+i\theta$. The superpotential can be written as \cite{Gukov:1999ya,Gukov:1999gr},
\begin{align}
    W& =W_0+W_{np}\nonumber \\
    &=W_0+A\,e^{-a\rho}\nonumber \\
    &=W_0+ A_0f(z)^{1/n}e^{-a\rho},
\end{align}
where $W_0$ takes care of the stabilization of the complex structure moduli and the dilaton. The addition of $W_{np}$ to $W$ constitutes the second step of the moduli stabilization. The prefactor $A_0$ depends on the complex structure moduli and the dilaton, while $A$ depends also on the open string moduli. $f(z)$ is a functional\footnote{As stated in appendix \ref{appA}, a conifold can be expressed in terms of two types complex coordinates. Hence, $f(z)$ can also be written as a function of $\omega$, constructing the Ouyang embedding \cite{Ouyang:2003df}.} that depends on the position of the brane and plays a particular role in the derivation of the brane potential as we shall see in the following sections. \par

In the next subsection, we shall investigate the structure of the NP effects  inside a WDC. The NP corrections are usually computed with the help of $D7$-branes, which then wrap certain divisors. The embedding of these divisors into the CY manifold, or especially into the non-compact conifold, will finally determine the structure of the NP effects of our interest. There are two large classes of such embeddings: Ouyang embedding \cite{Ouyang:2003df} and the Kuperstein embedding \cite{Kuperstein:2004hy}. In this paper, we focus on the latter class of supersymmetric embedding\footnote{It has been emphasized in \cite{Krause:2007jk} that only Kuperstein embedding enables the angular directions other than the inflaton to be put in their respective minima.} to generate the potential for both the radial and angular directions of our $D3$-brane.\par

\subsection{F-term scalar potential for the brane}

Throughout the work, the internal space will be represented by a WDC and the superpotential is given by~\cite{Baumann:2006th,Ganor:1996pe,Martucci:2006ij,Pajer:2008uy}
\begin{equation}\label{superpotential_gen}
    W=W_0+A_0f(z)^{1/n}e^{-a\rho}=W_0+A\,e^{-a\rho}.
\end{equation}
It is further shown in \cite{Krause:2007jk,Pajer:2008uy} that when one combines the K\"ahler potential of \eqref{kahler_pot_gen} and the superpotential of \eqref{superpotential_gen} in the F-term scalar potential \eqref{scalar_pot_gen}, one gets
\begin{equation}
    V_F=\frac{k_4^2}{3U^2}\left[ U|W_{,\rho}|^2-3(\bar{W}W_{,\rho}+\text{c.c.})+\frac{1}{\gamma}k^{\bar{i}j}\overline{W}_{,\bar{i}}W_{,j} \right],
\end{equation}
which can be further simplified as~\cite{Pajer:2008uy}
\begin{align}
    V_F=\frac{k_4^2}{3U^2}\left[ (Ua^2+6a)|A|^2e^{-a(\rho+\bar{\rho})} + 3 a(\bar{W}_0\,Ae^{-a\rho}+\text{c.c.}) +\frac{1}{\gamma}k^{\bar{i}j}\bar{A}_{\bar{i}}A_{,j}e^{-a(\rho+\bar{\rho})} \right].
\end{align}
This potential remains valid only for a $D3$-brane at or in the vicinity of the tip where the K\"ahler potential remains block diagonal \cite{Pajer:2008uy}. The above potential is a function of $(\sigma,x_1,x_2,x_3)$ (see appendix~\ref{appA} for details). As stated in \cite{Krause:2007jk}, addressing moduli stabilization leads to a potential that incorporates the KKLT \cite{Kachru:2003aw} as well as an extra piece taking care of the $D3$-brane part. The full potential reads~\cite{Krause:2007jk}
\begin{align}\label{fullpot}
    V_F&=V_{KKLT}+\Delta V,\\
    V_{KKLT}&= \frac{k_4^22a|A|e^{-a\sigma}}{U^2}\left(\frac{1}{6}aU|A|e^{-a\sigma}+|A|e^{-a\sigma}-|W_0|\right),\\
    \Delta V&=\frac{k_4^2e^{-2a\sigma}}{3U^2\gamma}k^{\bar{i}j}\overline{A}_{,\bar{i}}A_{,j}.
\end{align}

$V_{KKLT}$ is solely responsible for the K\"ahler moduli stabilization, $\Delta V$ takes care of the brane part, and the heavy closed string moduli, such as complex structure and axio-dilaton, are stabilized via the $W_0$ term in \eqref{fullpot}. When the superpotential  does not depend on the NP effects, the F-term potential for a $D3$-brane has a $SO(4)$ symmetry that is broken once we have chosen $W$ to take the form of \eqref{superpotential_gen}, or we have chosen a particular embedding. The final system can be effectively described in terms of a single real field, as will be discussed below.

\subsubsection{Kuperstein embeddings in angular inflation}

The $D3$-brane inflation at the tip occurs along one of the angular directions while being at a local minimum along the radial direction and the other $4$ angular directions of the entire Klebanov-Strassler (KS) throat~\cite{Klebanov:2000hb}. The $SO(4)$ symmetry of the KS throat is only broken by choosing a certain embedding function $f$. The embedding function that sources the NP effects is usually located on the divisor wrapped by the $D7$-brane. Of the three remaining directions, two remain as flat directions of the scalar potential due to the $SO(3)$ symmetry. Bulk contributions will eventually stabilize them, but the details of such contributions are beyond the scope of this paper. In our inflationary analysis, we will assume that the $D3$-brane in the warped throat stays at a minimum along the $x_2$ and $x_3$ directions and moves only along $x_1$. Next, following \cite{Pajer:2008uy}, we opt for a more convenient parametrization of $S^3$ such that
\begin{align*}
    & z_1=x_1=\varepsilon \cos\varphi, \hspace{2.6cm}z_2=x_2=\varepsilon \cos\varphi \sin \psi\sin\theta,\\
    & z_3=x_3=\varepsilon \sin\varphi \sin \psi\sin\theta, \hspace{1cm} z_4=x_4=\varepsilon \cos\theta\sin\psi.
\end{align*}
This leads to a diagonal metric of the form,
\begin{equation}
    ds^2=c\,\varepsilon ^{4/3}\left[d\varphi^2+\sin^2\varphi(d\theta^2+\sin^2\theta d\psi^2)\right].
\end{equation}
Under Kuperstein embedding \cite{Kuperstein:2004hy}, the NP superpotential takes the form,
\begin{equation}\label{Kuperstein_NP}
    W_{np}=A_0\left(1-\frac{x_1}{\mu}\right)^{1/n}e^{-a\rho},
\end{equation}
where in the above, we choose a case where the $D7$-brane of our interest does not go near the tip, hence $\varepsilon\ll \mu$. Angular inflation occurs along the $\varphi$-direction. In the presence of the NP term like \eqref{Kuperstein_NP}, the KKLT potential and the extra part that goes into the formation of the potential for the $D3$-brane become
\begin{align}\label{Angular_brane_pot1}
    V_{KKLT}\simeq &~ \frac{2k_4^2|A_0|a e^{-a\sigma}}{U^2}\left(\frac{1}{6}a|A_0|Ue^{-a\sigma}+|A_0|e^{-a\sigma}-|W_0|\right)\nonumber\\
    &~ +\frac{2k_4^2\varepsilon|A_0|ae^{-a\sigma}}{U^2n\mu}\left(\frac{1}{3}|A_0|aUe^{-a\sigma}+2|A_0|e^{-a\sigma}-|W_0|\right)\cos\varphi+\nonumber\\
    &~ + \frac{2k_4^2|A_0|ae^{-a\sigma}\varepsilon^2}{U^2n\mu^2}\bigg(2|A_0|e^{-a\sigma}-\frac{3|A_0|e^{-a\sigma}}{n}+\frac{1}{3}a|A_0|Ue^{-a\sigma}-\frac{a|A_0|Ue^{-a\sigma}}{2n}\nonumber\\
    &~ -|W_0|\left(\frac{n-1}{n}\right)\bigg)\cos^2\varphi+...\;,\\
    \Delta V\simeq &~ \frac{k_4^2|A_0|^2e^{-2a\sigma}\varepsilon^{2/3}}{3cn^2U^2\gamma \mu^2}\sin^2\varphi-\frac{2k_4^2|A_0|^2e^{-2a\sigma}\varepsilon^{5/3}}{3cn^3U^2\gamma \mu^3}\sin^2\varphi\cos\varphi.
\end{align}
In the potential written above, the first line provides the dominant contribution and being the standard KKLT term. Without a proper uplift term, it gives rise to an AdS minimum for the K\"ahler modulus $\sigma$. The other terms are suppressed by orders of $\varepsilon/\mu$, but still play a pivotal role in driving warm $D3$-brane inflation, as we shall discover soon. Contrary to \cite{Pajer:2008uy}, in our work we do consider the expansion in $\varepsilon\ll\mu$, but keep one more sub-leading term in the series. The reason is that, as demonstrated in section \ref{sec:4}, it leads to observationally viable warm inflationary models. Following \cite{Pajer:2008uy}, we remain agnostic about the true origin of the uplift term as there are wide ranges of options such as D-term uplift \cite{Burgess:2003ic}, the T-brane uplift \cite{Cicoli:2015ylx}, F-term uplift \cite{Brax:2007xq} or an anti-brane uplift \cite{Kachru:2002sk,Crino:2020qwk,Bento:2021nbb}. The radial stability of the D-brane is extensively carried out in the appendices of \cite{Pajer:2008uy} and since our potential changes from \cite{Pajer:2008uy} only at the sub-leading order, we do not expect any substantial change to the stabilization mechanism of the radial direction. \par
The potential of \eqref{Angular_brane_pot1} can be substantially simplified  and can be written only in terms of the inflaton $\varphi$, as the K\"ahler modulus $\sigma$ is stabilized by the leading order terms in the first line \`a la KKLT \cite{Kachru:2003aw}. Therefore, the potential takes the following form,
\begin{align}\label{Angular_brane_pot2}
    V(\langle\sigma\rangle,\varphi)=&~V_{up}+V_{KKLT}+\Delta V,\nonumber\\
=&~\Lambda(\sigma)+A(\sigma)\cos\frac{\varphi}{d}+B(\sigma)\cos^2\frac{\varphi}{d}+\left(C(\sigma)-D(\sigma)\cos\frac{\varphi}{d}\right)\sin^2\frac{\varphi}{d}.
\end{align}
Two comments are in order: first, $\Lambda(\sigma)$ takes care of the K\"ahler modulus dependent uplift term. $\sigma$ is stabilized at the standard KKLT minimum at
\begin{equation}\label{KKLT_stab}
    W_0=-A_0e^{-a\langle\sigma\rangle}\left(\frac{2}{3}a\langle\sigma\rangle+1\right).
\end{equation}
Second, the scalar $\varphi$ is the canonically normalized inflaton with $\varphi=\varepsilon^{2/3}\sqrt{T_{D3} c}\, \tilde{\varphi}=d\, \tilde{\varphi}$. Note that $\varphi$ has mass dimension. The coefficients $\{A(\sigma),B(\sigma),C(\sigma),D(\sigma)\}$ can be read by directly comparing the terms between \eqref{Angular_brane_pot1} and \eqref{Angular_brane_pot2}.\par

For the sake of studying a successful model of slow-roll, we will further simplify \eqref{Angular_brane_pot1} to:
\begin{equation}
    V(\varphi)=\Lambda_1\left(1+\alpha_1\cos\frac{\varphi}{d}+\alpha_2\cos^2\frac{\varphi}{d}\right)+\Lambda_2\left(1-\beta\cos\frac{\varphi}{d}\right)\sin^2\frac{\varphi}{d}.
\label{pot-angular}
\end{equation}
where
\begin{align*}
    \Lambda_1=&~\Lambda(\sigma),\quad
    \alpha_1=~\frac{A(\sigma)}{\Lambda_1(\sigma)},\quad
    \alpha_2=~\frac{B(\sigma)}{\Lambda_1(\sigma)},\quad \Lambda_2=C(\sigma),\quad \beta=\frac{D(\sigma)}{C(\sigma)}.
\end{align*}
\vspace{0.5em}

\subsubsection{Kuperstein embeddings in radial inflation}

The standard radial brane inflation scenario was studied in \cite{Krause:2007jk}. In contrast to earlier brane–anti-brane inflation models, and similar to the case of angular brane inflation, this setup also incorporates the effects of moduli stabilization. The brane potential is induced by the threshold corrections to the NP superpotential. The radial coordinate dependent potential is derived after the angular moduli are stabilized. The scalar potential of the original brane inflation \cite{Kachru:2003sx} features a Coulomb interaction, developed due to the interactions between a brane and an anti-brane. In this work, we consider both the effects from the NP superpotential as well as a sub-leading Coulomb interaction term sourced via brane and an anti-brane (if present at the tip but not coincident with the $D3$-brane). In this class of embedding, the NP effects take the following form:
\begin{equation}
    A=A_0f^{1/n}=A_0\left(1+\frac{r^{3/2}}{\mu}\right)^{1/n}\simeq A_0\left(1+\frac{r^{3/2}}{\mu n}\right).
\end{equation}
Similarly to the angular brane inflation case, the total scalar potential can be categorized into a KKLT part and the rest taking care of the brane potential. Following \cite{Krause:2007jk}, the potential in the limit $r^{3/2}\ll \mu$ can be written as
\begin{align}\label{radial_brane_pot1}
   & V(\sigma,r) = ~V_{KKLT}+\Delta V,\\
  &  V_{KKLT}~\propto V_{KKLT}^0 \propto f(r^2),\\
  &  \Delta V = ~ \frac{k_4^2|A_0|^2e^{-2a\sigma}}{n^2U^2}\left[\frac{2\pi r^{3/2}}{\sqrt{2}\mu+r^{3/2}}+\frac{r}{\gamma(\sqrt{2}\mu+r^{3/2})^2}\right],
\end{align}
where $V_{KKLT}^0$ denotes the original KKLT potential without the uplift term. The potential driving the radial brane inflation can be written in a simplified manner as\footnote{We refer to appendix D of \cite{Krause:2007jk} to understand the negative sign of $r^{3/2}$ term, and we intend to not repeat the calculation here.},
\begin{equation}
    V(\phi)=V_0\left(\Lambda+C_1\phi-C_{3/2}\phi^{3/2}+C_2\phi^2-\frac{D}{\phi^4}\right),
\label{pot-radial}
\end{equation}
where, similar to angular brane inflation, $\phi$ denotes the canonical version of the original radial field and is related by $\phi=\sqrt{T_{D3}}r$. $\Lambda$ takes care of the uplift term, and $\{C_1,C_{3/2}\}$ can be read from \eqref{radial_brane_pot1}. $C_2$ depends on the KKLT potential with a K\"ahler modulus stabilized as in \eqref{KKLT_stab}. $D$ is the Coulomb term that depends on the tension of the $D3$-brane $T_{D3}$, the warping of the throat depending on the type-IIB fluxes, as well as the characteristic length scale of the warped throat. \par

Next, we analyze the inflationary dynamics arising from the potentials given in eqs.~(\ref{pot-angular}) and (\ref{pot-radial}) for the two brane model realizations discussed above.

\section{Warm inflation}\label{sec:3}

Since the main goal of our paper is to embed brane inflation in WI, let us briefly review the basics of the WI dynamics (see \eg~\cite{Kamali:2023lzq} for a recent review) in order to set the stage for the results displayed in this paper to be presented in the next section.

\subsection{Background dynamics}

Previous CI models explored in the context of angular and radial brane inflation typically lead to observationally excluded models when compared with current and forthcoming data. We offer a remedy to such drawbacks by successfully embedding them in WI, which facilitates a successful period of slow-roll inflation viable with observations. This comes about due to the dissipative effects inherent of the WI dynamics and the coupling to the radiation bath generated and maintained during WI. 

In WI, the inflaton field $\phi$ is coupled to the dynamics of the radiation energy density $(\rho_r)$ of the thermal bath. The presence of a radiation bath introduces a dissipation coefficient $(\Upsilon)$, such that the background equations for the inflaton field and the radiation energy density become 
\begin{align}
   & \Ddot{\phi}+(3H+\Upsilon)\dot{\phi}+V_{,\phi}=0,
   \label{eq1}\\
   & \dot{\rho}_r+4H\rho_r=\Upsilon \dot{\phi}^2,
   \label{eq2}
\end{align}
where $V_{,\phi}=\frac{d V(\phi)}{d\phi},$ and $H$ is the Hubble parameter, 
\begin{equation}
    H^2=\frac{1}{3M_{\rm Pl}^2}\left(\frac{\dot{\phi}^2}{2}+V+\rho_r\right).
    \label{eq3}
\end{equation}
In the presence of a thermalized radiation bath, $\rho_r=C_RT^4$, with $C_R=\pi^2 g_{\star}/30$ and $g_{\star}$ is the effective number of degrees of freedom for the radiation, e.g. $g_*=106.75$ for the case of the SM of particles, or $g_*=228.75$ for the minimal supersymmetric SM (MSSM). In all of our numerical examples we choose the $g_*$ like for the MSSM. However, the results are weakly dependent on the chosen value for $g_*$, e.g. when computing the spectral tilt of the scalar of curvature power spectrum or background quantities. Similarly to CI, in WI, the end of inflation is signaled by $\epsilon_H=1$, where
\begin{equation}
    \epsilon_H=-\frac{\dot{H}}{H^2}\simeq \frac{\epsilon_V}{1+Q},\qquad \epsilon_V=\frac{M_{\rm Pl}^2}{2}\left(\frac{V_{,\phi}}{V}\right)^2,
\label{eq4}
\end{equation}
and $Q=\Upsilon/(3H)$ denotes the dissipation ratio in WI. $Q$ determines the amount of energy dissipated to the radiation bath during inflation compared to the Hubble parameter $H$. $Q\gtrsim 1$ refers to the strong dissipative regime and $Q\ll 1$  to the weak dissipative regime in WI. 

\subsection{Perturbation in WI and dissipation mechanisms}

The primordial scalar curvature power spectrum for a WI models can be expressed as~\cite{Ramos:2013nsa,Kamali:2023lzq}
\begin{equation}\label{power_spectra}
    P_{R}\simeq \left(\frac{H^2}{2\pi\dot{\phi}}\right)^2\left(1+2n_{\star}+\frac{2\sqrt{3}Q}{\sqrt{3+4\pi Q}}\frac{T}{H}\right)G(Q)\big|_{k_{\star}=a_\star H_\star},
\end{equation}
where $n_\star$ equals to $n_{\star}=n_{BE}\equiv 1/(e^{H/T}-1)$ in the case of thermalization of the inflaton perturbations and $n_{\star}=0$ in the absence of thermalization of the inflaton perturbations. The function $G(Q)$ in \eqref{power_spectra} depends mainly on the dissipation coefficient~\cite{Montefalcone:2023pvh}, but a more recent analysis~\cite{Rodrigues:2025neh} indicates that it can also exhibit some dependence on the model when $n_{\star}=0$. The general expression for the dissipation coefficient can be expressed as \cite{Bastero-Gil:2010dgy,Bastero-Gil:2012akf,Bastero-Gil:2016qru}:
\begin{equation}\label{Upsilon}
    \Upsilon=C_{\Upsilon}f(T,\phi),\qquad f(T,\phi)=T^c\phi^pM_{\rm Pl}^{1-p-c},
\end{equation}
where $C_{\Upsilon}$ is a dimensionless constant, which depends on the microscopic details of the model, i.e. the coupling between the thermal bath and the inflaton. Using \eqref{power_spectra}, the scalar spectral index $n_s$ can be calculated at the Hubble exit point $(k_{\star})$,
\begin{equation}
    n_s-1=\frac{d\ln\,P_{\mathcal{R}}(k)}{d\ln\,k}\bigg|_{k\to k_{\star}},
\end{equation}
while the tensor-to-scalar ratio $r$ is given by
\begin{align}
    r&=\frac{P_{\mathcal{T}}}{P_{\mathcal{R}}}, 
\end{align}
where $P_{\mathcal{T}}$ is the tensor spectrum,
\begin{equation}
P_{\mathcal{T}}=\frac{2H^2}{\pi^2M_{\rm Pl}^2}.
\end{equation}

One of the striking features of WI is that in principle there is no need for a posterior reheating epoch. This eliminates, for instance,  the uncertainty appearing in CI studies in the number of e-folds between the horizon exit of relevant scales $(k_{\star})$ and their reentry. This allows us to directly calculate the appropriate $N_\star$ relevant for obtaining the cosmological observables (see, e.g.~\cite{Das:2020xmh,Rodrigues:2025neh}),
\begin{equation}\label{efold}
    \frac{k_{\star}}{a_0H_0}=e^{-N_{\star}}\left[\frac{43}{11g_s(T_{end})}\right]^{1/3}\frac{a_{end}}{a_{reh}}\frac{T_0}{T_{end}}\frac{H_{\star}}{H_0},
\end{equation}
where this is obtained by applying the entropy conservation results and equating $k_{\star}=a_{\star}H_{\star}$ with $k_0=a_0H_0$ with $0$ being the present time.  In eq. \eqref{efold}, $g_s(T_{end})$ denotes the entropy number of degrees of freedom at the end of inflation. For the exact computation, we use the following set of parameters: $k_{\star}=0.05\,\mathrm{Mpc^{-1}}$, the value of the scale factor today is assumed to be $a_0=1$, the Hubble parameter and the temperature of cosmological microwave background (CMB) today given, respectively, by~\cite{Planck:2018vyg} $H_0=67.66\,\mathrm{km\,s^{-1}Mpc^{-1}}$ and $T_0=2.725\,\mathrm{K}\simeq 2.349\times 10^{-13}\,\mathrm{GeV}$.
In all of our numerical results, we make use of the
Mathematica package \texttt{WI2easy} code~\cite{Rodrigues:2025neh} to obtain the background dynamics and the perturbations for the different models studied here. \texttt{WI2easy} follows a deterministic {}Fokker-Planck method for the numerical evolution of perturbations in WI~\cite{Ballesteros:2022hjk,Ballesteros:2023dno}.

\subsection{The form of the dissipation function}

The dissipation coefficient in WI introduced in \eqref{Upsilon} depends in a very intricate way on the microphysics of the underlying model (for reviews, see, e.g.,~\cite{Berera:2008ar,Kamali:2023lzq}). For instance, in this paper, we study angular inflation, where the inflaton is an axion-like particle, and in the case of radial brane inflation, it is a saxion. For axions, the most well-known dissipation mechanism is when the inflaton dissipates its energy to radiation gauge bosons through sphaleron processes in a thermal bath~\cite{Berghaus:2019whh}. {}For saxions, one of the motivated scenarios is the two-stage dissipation mechanism~\cite{Berera:2008ar} and further studied in~\cite{Bastero-Gil:2010dgy,Bastero-Gil:2012akf}. {}For the case of radial brane inflation, we opt for the similar dissipation mechanism considered in~\cite{Bastero-Gil:2011zxb}, where the inflaton is associated with the distance between the UV end and the IR end of a warped throat (see figure~\ref{fig:illustration1}). 
{}For angular brane inflation, where the inflaton can be naturally coupled to gauge fields with a dimension five axion-like form, we consider the dissipation coefficient as in minimal WI~\cite{Berghaus:2019whh}. In both cases mentioned above, the inflaton potential is protected against large quantum and thermal corrections, either due to the shift symmetry of axion-like inflaton field, or in the case of the radial inflation case, we consider that the inflaton is only coupled
to heavy fields, with, hence, Boltzmann suppressed thermal corrections and quantum corrections suppressed due to (approximated) SUSY.
Below, we motivate the form of dissipation for these two cases of warm brane inflation: the angular and radial ones.

\subsubsection{Angular WI}

In the angular WI case, we consider the inflaton to be an angular coordinate of a $S^3$ sphere at the tip of a WDC. Being an angular direction, it essentially acts like an axion with shift-symmetry, which helps in maintaining the flatness of the inflaton potential against quantum and thermal corrections. Now, to understand the interaction between the inflaton and the gauge sector, let us write down the following interaction Lagrangian density in terms of the gauge fields and a complex scalar field, $\Phi=\varepsilon e^{i\varphi/d}$,
\begin{equation}
    \mathcal{L}_{int}\simeq -\frac{f(\Phi)}{4}F_{\mu\nu}^AF^{A\,\mu\nu}+\frac{h(\Phi)}{4}F_{\mu\nu}^A\tilde{F}^{A\,\mu\nu},
\label{Lintaxion}
\end{equation}
where the inflaton corresponds to the phase $\varphi$ and $F^A_{\mu\nu}$ denotes the gauge field strength. The functions $h$ and $f$ in (\ref{Lintaxion}) dictate the coupling between the gauge sector and the inflaton and is dependent on the scalar fields. In our setup, the K\"ahler moduli are stabilized through the KKLT mechanism, and only the brane coordinates drive WI. The ref.~\cite{Holland:2020jdh} studied a K\"ahler moduli driven multi-field inflationary model. In that reference, the SM was realized in a magnetized brane and, hence, their coupling functions become K\"ahler moduli dependent. The saxionic part of the K\"ahler modulus determines the coupling to the gauge kinetic term $F^2$, while the axionic part determines the coupling to the term $F\tilde{F}$. 

In our case, however, the coupling arises from our complex scalar $\Phi$, where the axion $\varphi$ determines the function $h$, while and the modulus of the field, $\varepsilon$, determines $f$. Hence, to study angular brane inflation, we consider the radius of the $S^3$ sphere \ie, $\varepsilon$ to be stabilized by type-IIB fluxes \cite{Bento:2021nbb} and integrated out. In this context, $f(\Phi)=\varepsilon/M_{\rm Pl}$ and $h(\Phi)=n\varphi/d$ where $n$ is the rank of the gauge group and $d$ is the effective decay constant of the axion defined below \eqref{KKLT_stab}. 

Now, if we consider the inflaton $\varphi$ to be coupled to an arbitrary Yang-Mills group with gauge coupling $g_{YM}$, as shown in \cite{Berghaus:2019whh}, this type of interaction leads to a friction coefficient $\Upsilon$ being proportional to $T^3$. In the presence of non-chiral fermions, the dissipation can become $\propto T$ 
instead~\cite{Berghaus:2020ekh,Drewes:2023khq,Berghaus:2024zfg}. Superstring theory will naturally allow for the presence of such fermionic components. 
Here we notice that inflaton potentials of standard cosine-like form,
including the one considered here, eq.~(\ref{Angular_brane_pot2}), when considered in WI with a dissipation coefficient $\propto T^3$, are only consistent with the observations for super-Planckian axion decay constants~\cite{Montefalcone:2022jfw,Zell:2024vfn,Bastero-Gil:2026ypn}. On the other hand, as shown here, 
sub-Planckian axion decay constants are allowed in the case of
a dissipation coefficient $\propto T$, hence better suited for a string-motivated effective model standpoint as studied in this work.
Therefore, for our angular brane part, we specifically use the dissipation function with a linear temperature dependence, $\Upsilon \propto T$.

\subsubsection{Radial WI}

While angular brane inflation takes place in the $S^3$ of the WDC, radial brane occurs in a direction orthogonal to it along the length of the throat. For radial brane inflation, inflaton is the radial coordinate of a deformed conifold. It is also the separation modulus between $D3-D7$. In this case, the dissipation mechanism is motivated by the two-stage dissipation mechanism first described in \cite{Bastero-Gil:2010dgy}. In the model realization motivating this dissipation mechanism, the inflaton is coupled to the heavy fields, while the heavy fields are further coupled to light radiation fields, through which they can decay and the energy of the inflaton be transferred to. These lighter fields do not directly couple to the inflaton sector. 

For the model realization of the above interaction scheme, we follow, for instance, ref.~\cite{Bastero-Gil:2011zxb}. As described in \cite{Bastero-Gil:2011zxb}, to understand the interaction between inflaton and the SM sector, let us define the following fields: if we have two D3 branes (say A and B) in the system, then their position in the compact space denotes fields expressed as $(\Phi_2,\Phi_3)$, whereas the distance between the D3-D7 brane is denoted by $\Phi_1$. The heavy fields are denoted by $X$, which are the strings stretched between the D3-branes and the light fields $Y$, which denote the strings stretched between $D3-D7$. Then according to \cite{Bastero-Gil:2011zxb}, the interaction superpotential can be expressed in the form
\begin{equation}
W=\sqrt{2}g_{YM}\left[\Phi_1^{AB}X_2\tilde{X}_3+\tilde{Y}^B\tilde{X}_3Y^A+...\right],
\label{model1}
\end{equation}
where the scalar component of the superfield $\Phi_1$,
$\phi$, is the inflaton candidate. Although superfields $\Phi_2$ and $\Phi_3$ are coupled to heavy fields $X$, they can indirectly transfer their energy to light fields $Y$ through the decay of intermediate heavy $X$ to $Y$. Since the $Y$ superfields are coupled to the non-inflaton sector, they can remain light when the brane resides close to the tip of the warped throat. This is how radial brane inflation can justify having a two-stage dissipation mechanism and friction coefficient of the form $\Upsilon \propto T^3/\phi^2$ (see, e.g. refs.~\cite{Bastero-Gil:2010dgy,Bastero-Gil:2012akf}).

\section{Numerical results}\label{sec:4}

Let us now present our numerical analysis for each of the two types of brane inflation
discussed in the previous sections. All numerical computations are performed using the publicly available Mathematica package \texttt{WI2easy} code for WI analysis~\cite{Rodrigues:2025neh}.
Let us first start with the case of radial brane WI.

\subsection{Radial brane WI results}

The potential given in \eqref{pot-radial} describes the motion of a mobile D3-brane along the radial direction of a WDC. Contrary to Ouyang, Kuperstein embedding allows us to put the angular directions at their minima. As shown in eq. \eqref{radial_brane_pot1}, the $\Delta V$ term, \ie, the embedding effect is reflected on the $\phi$ and  $\phi^{3/2}$ terms, whereas the effect of moduli stabilization \`a la KKLT and the uplift can be seen in the terms proportional to $\phi^2$ and in the constant term. Finally, we have also incorporated a sufficiently sub-leading brane-anti-brane interaction, \ie, the Coulomb potential, which has no effect in the extremization of the potential. 

\begin{table}[H]
\caption { Model parameters (in units of $M_{\rm Pl}$) for the potential \eqref{pot-radial}.}
\begin{center}
\centering
    \resizebox{0.8\textwidth}{!}{ 
    \begin{tabular}{| l | c | c | c | c | }
\hline
\cellcolor[gray]{0.9} $\Lambda$ &\cellcolor[gray]{0.9} $C_1$ &  \cellcolor[gray]{0.9} $C_{3/2}$ &  \cellcolor[gray]{0.9} $C_2$ &  \cellcolor[gray]{0.9} $D$  \\
\hline \hline
 $6.2\times 10^{-5}$ & $1.57\times 10^{-4}$  & $2.4\times 10^{-4}$ & $8\times 10^{-5}$ & $10^{-8}$\\
\hline
\end{tabular}}
\end{center} 
\label{tab1}
\end{table}

\begin{figure}[!htb]
    \centering
    \includegraphics[width=0.8\linewidth]{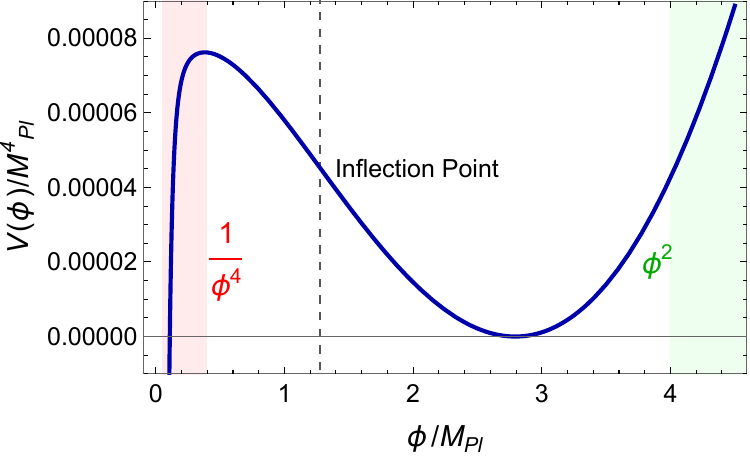}
    \caption{The radial-brane potential \eqref{pot-radial} for the parameters given in table \ref{tab1}}
    \label{fig:illustration1_radial_brane}
\end{figure}

In table~\ref{tab1}, we give a set of representative constants in the potential~\eqref{pot-radial} and which we will use here as a benchmark model for our numerical analysis. The corresponding potential for those parameters is shown in figure~\ref{fig:illustration1_radial_brane}.
This choice of parameters generates a maximum, a minimum, and a point of inflection for the potential. The negative asymptote appears for small values of the inflaton dominated by the Coulomb term. The potential displays a $\phi^2$ type behavior at large $\phi$. In the original work \cite{Krause:2007jk}, it is claimed that the potential can only be made flat enough when the distinction between a maximum and a minimum disappears, further flattening the inflection point. However, as we shall show in this work, with our parameter choice, one can realize a successful inflationary model both in the cold and warm regime despite not having a flat inflection point region. We also justify the choice of parameters to their counterpart in string theory.

As can be easily checked for the potential of the form \eqref{pot-radial}, the existence of extrema is dictated by demanding that $9C_{3/2}-32C_1C_2>0$. For the benchmark parameters chosen, we can satisfy this inequality and also find a slow-roll inflation regime. As expressed in \cite{Krause:2007jk}, the coefficients can be mapped in the string compactification and the $D3$-brane parameters as follows:
\begin{align}
    \Lambda & \propto \frac{a^2}{6\sigma},\;\; C_1\propto \frac{1}{T^{3/2}_{D3}\sigma^{3/2}\mu^2n^2},\;\;
    C_{3/2} & \propto \frac{a }{12 \sigma^{5/4}T_{D3}^{3/4}\mu n},\;\; C_2<\Lambda, \;\; D\propto T_{D3}h_0^4R^4.
\end{align}

$C_2$ is the uplifting term and can therefore be arbitrarily chosen to be small. $\mu^{2/3}$ is the distance between $D3-D7$, $a=2\pi\,\mathcal{O}(1/n)$ and $\sigma$ here is the stabilized value of the K\"ahler modulus following the KKLT mechanism. $T_{D3}$ is the tension of the $D3$-brane, which is given by $T_{D3}=1/(2\pi g_s l_s^4)$. For the parameter chosen in table \ref{tab1}, we can set $\sigma\sim\mathcal{O}(100)$, string length $l_s\sim\mathcal{O}(10)$, residing in the large rank stack of $D7-$branes (building explicit models) \cite{Louis:2012nb,Carta:2019rhx}, and string coupling $g_s\ll 1$. We also have that the hierarchy between the coefficients is: $  \{C_1,C_{3/2}\}>C_2>\Lambda$ and $C_{3/2}>C_1$, while keeping the discriminant greater than zero. $D$ is sub-dominant as it is proportional not only to $T_{D3}$ but also to $h_0$, which is the warping at the tip of a WDC. This is chosen to be small in order for the brane-energy to not overshoot moduli stabilization and boost mild uplift of the AdS minima. The overall factor $V_0$ can be chosen by giving suitable values for $W_0$ and $A_0$ in \eqref{superpotential_gen}.       

\begin{center}
\begin{figure}[!bth]
\subfigure[]{\includegraphics[width=7.5cm]{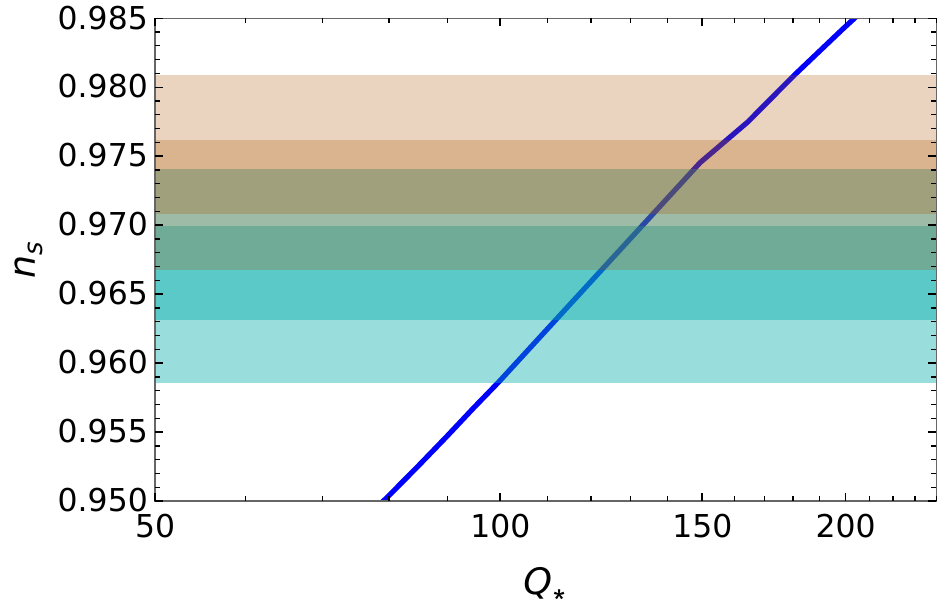}}
\subfigure[]{\includegraphics[width=7.5cm]{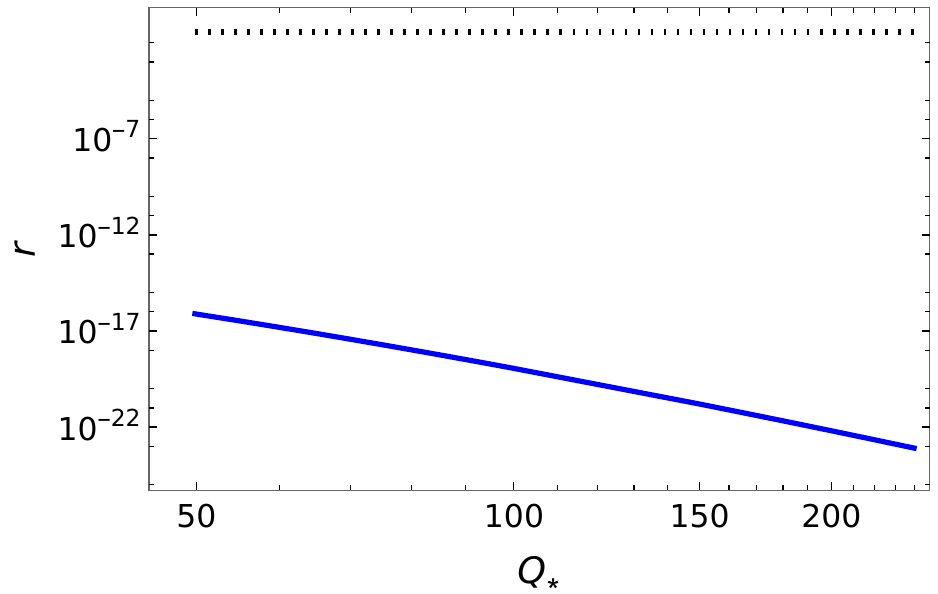}}
\caption{Plot of $n_s$ (panel a) and $r$ (panel b) as  a function of $Q_{\star}$ (at the Hubble radius crossing point) for the case of the radial direction brane inflation. The horizontal dotted line shown in panel (b) indicates the upper bound $r< 0.036$, from ref.~\cite{BICEP:2021xfz}.}    
\label{fig:4}
\end{figure}
\end{center}

\begin{figure}[!htb]
    \centering
    \includegraphics[width=0.63\linewidth]{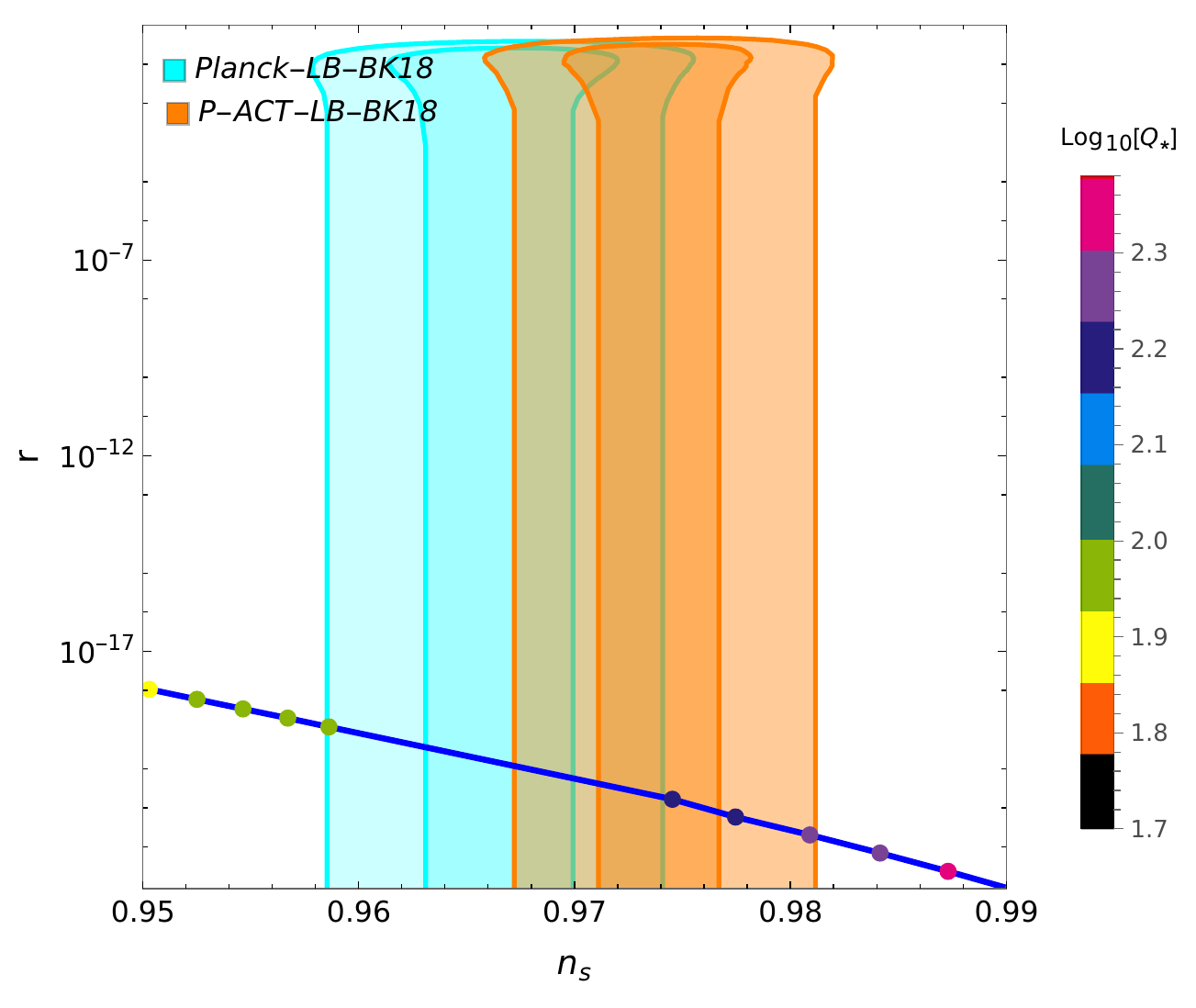}
    \caption{Plot of $n_s-r$ against Planck and ACT data for the case of the radial direction brane model in WI with dissipation coefficient $\Upsilon \propto T^3/\phi^2$. }
    \label{fig:5}
\end{figure}

\begin{table}[H]
\caption { Predictions from both CI and WI for the parameters of table~\ref{tab1} for the radial brane inflation with dissipation
coefficient $\Upsilon= C_{T^3} T^3/\phi^2$.}
\begin{center}
\centering
    \resizebox{0.8\textwidth}{!}{ 
    \begin{tabular}{| l |  c | c | c | c | c |}
\hline
 \cellcolor[gray]{0.9} $V_0/M_{\rm Pl}^4$ & \cellcolor[gray]{0.9} $Q_{\star}$ &  \cellcolor[gray]{0.9} $n_s$ &  \cellcolor[gray]{0.9} $r$ &  \cellcolor[gray]{0.9} $|\Delta \phi|/M_{\rm Pl}$  & \cellcolor[gray]{0.9} $C_{T^3}$\\
\hline \hline

 $4.90\times 10^{-24}$ & $140.0$  & $0.972$ & $3.6\times 10^{-21}$ & $0.98$   & $9.17\times 10^{12}$\\
\hline
\end{tabular}}
\end{center} 
\label{tab2}
\end{table}
In table~\ref{tab2}, we present the results for both cold and WI for the parameters given in table~\ref{tab1}. The dynamics and perturbations are evaluated assuming $n_\star=n_{BE}$ in the power spectrum for WI, Eq.~(\ref{power_spectra}) and which is expected to be appropriate for $Q_\star>1$.
For the parameters of table \ref{tab1} we have found a suitable region on the left-hand side of the minimum, 
\ie, between the maximum and the minimum with an initial condition close to the hilltop. The inflationary trajectory feels the effect of all the terms of different powers of $\phi$ except the Coulomb term, as expected. For CI on the other hand, sub-Planckian field excursion can only be achieved by localizing the field very close to the maximum. With the present model parameters CI also produces a too red-tilted value for the spectral index, thus not satisfying recent results for these cosmological quantities~\cite{Planck:2018jri,BICEP:2021xfz,BICEPKeck:2024stm,ACT:2025tim,ACT:2025fju}. As in this example successful inflation with enough e-folds is a highly fine-tuned situation, we refrain to show the results for CI for this model here. 
On the other hand, all these problems are mitigated in WI in the strong dissipative regime. For the chosen radial brane inflation, for $Q\gtrsim 140$ the field excursion is always sub-Planckian, 
$\Delta \phi \lesssim M_{\rm Pl}$, there is also no need for fine-tuned initial conditions since the extra friction due to the dissipation allows the inflaton to be away from the maximum position of the potential. Finally, as shown by the results in table~\ref{tab2}, we get a not too red tilted spectrum, consistent with the latest data (see figs.~\ref{fig:4} and \ref{fig:5}).
The adoption of our specific set of benchmark parameters is motivated by a clear difference in the results between CI and WI. This provides us with one of the many motivations to study radial brane inflation in the context of WI.

We note here that from the two-stage warm inflation model given by eq.~(\ref{model1}), the dissipation coefficient constant $C_{T^3}$ in terms of the coupling constant $g_{YM}$ and
when taking into account plasma masses for the fields and their decay widths, is given by~\cite{Bastero-Gil:2012akf}
\begin{equation}
C_{T^3} \simeq 0.16 g_{YM}^2 N_Y,
\label{CT3}
\end{equation}
where $N_Y$ is the number of decay channels for the light radiation fields $Y$. This then implies that in general we need a large number of light fields for this model, as it is characteristic for the form of dissipation coefficient originating from eq.~(\ref{model1}) (see, e.g., also ref.~\cite{Bartrum:2013fia}).

As is clear from table~\ref{tab2}, the presence of dissipation enormously helps the radial brane inflation model studied in \cite{Krause:2007jk} to fall into the observation window. Also note that since~\cite{Motaharfar:2018zyb} $\Delta \phi_{\text{warm}}\sim\Delta\phi_{\text{cold}}/(1+Q)$, then when $Q > 1$ dissipation actually helps to reduce the magnitude of the field excursion, bringing WI more in line with the distance conjecture in string theory.

\begin{center}
\begin{figure}[!htb]
\subfigure[]{\includegraphics[width=4.7cm]{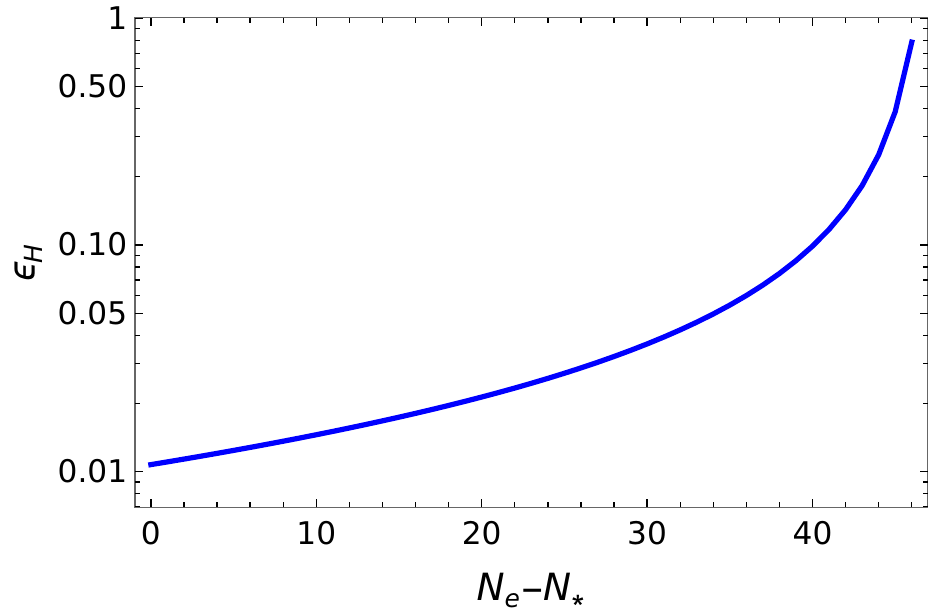}}
\subfigure[]{\includegraphics[width=4.6cm]{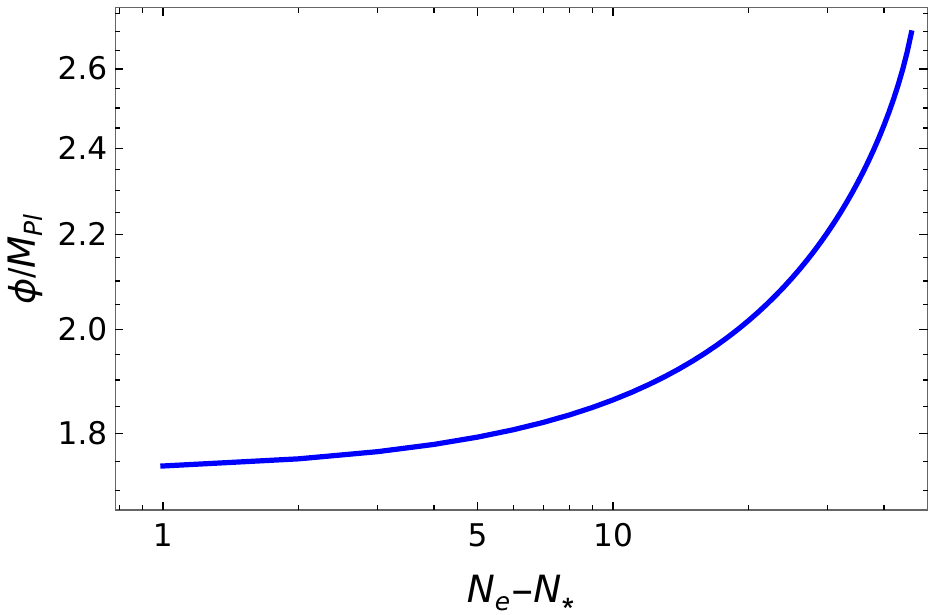}}
\subfigure[]{\includegraphics[width=4.8cm]{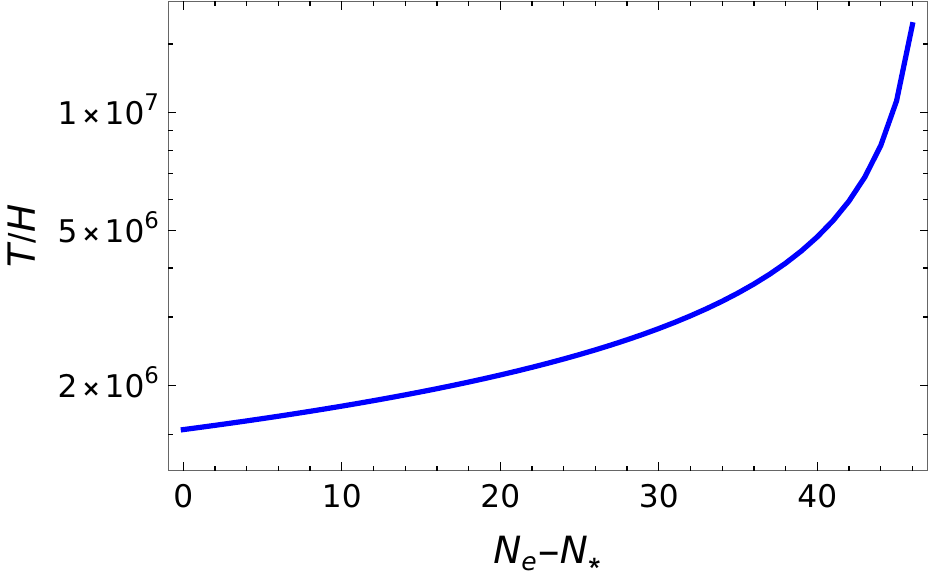}}
\subfigure[]{\includegraphics[width=4.8cm]{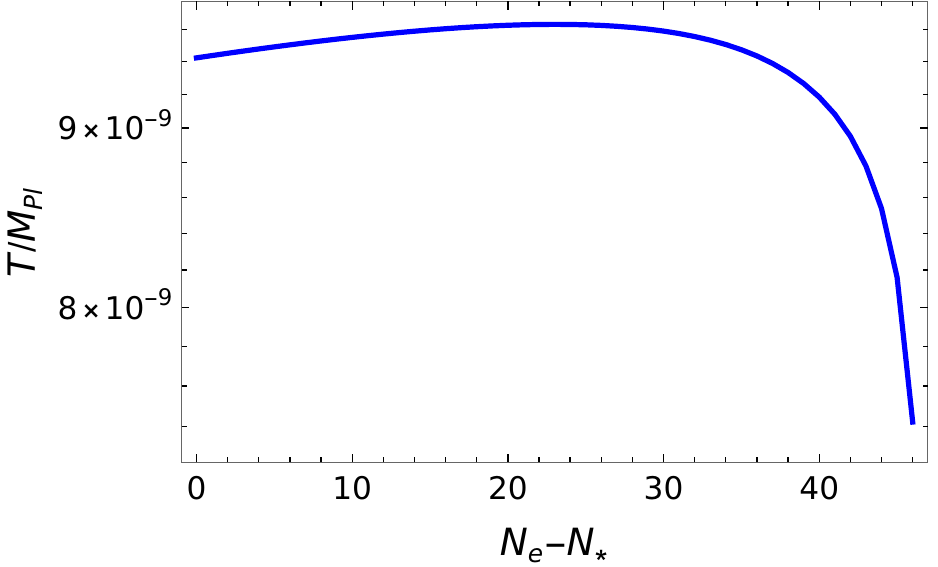}}
\subfigure[]{\includegraphics[width=4.6cm]{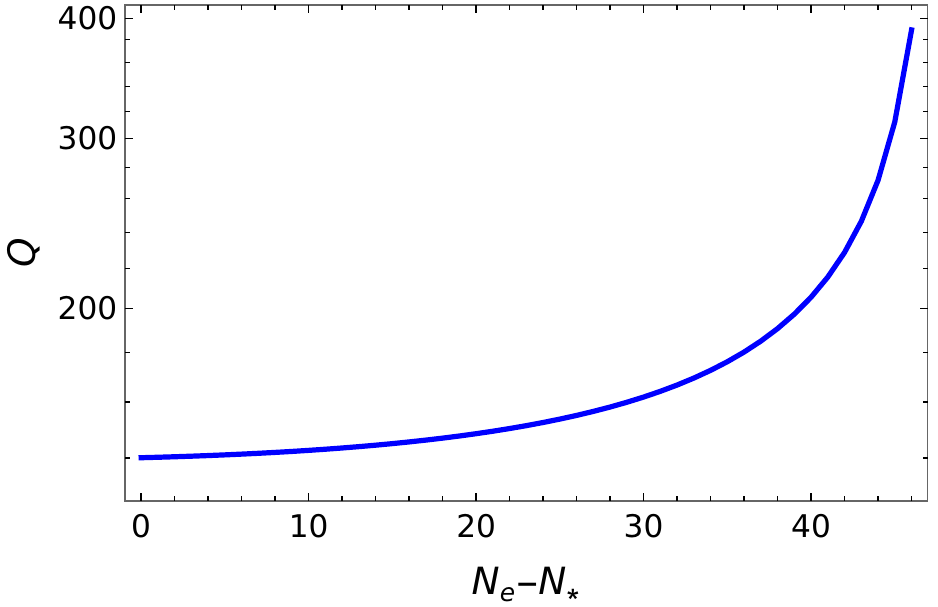}}
\hspace{0.5cm}
\subfigure[]{\includegraphics[width=4.6cm]{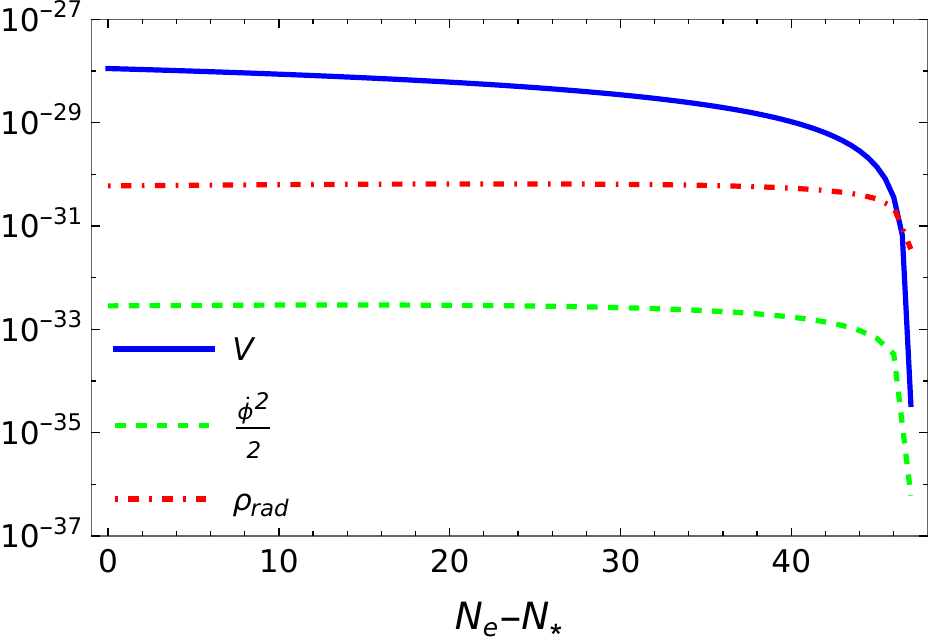}}
\caption{Evolution for the slow-roll coefficient $\epsilon_H$ (panel a), inflaton amplitude $\phi$ (panel b), $T/H$ (panel c), $T$ (panel d), $Q$ (panel e) and energy densities (panel f), in units of the reduced Planck mass
$M_{\rm Pl}$, for the radial brane WI model given in table~\ref{tab2}.}    
\label{fig:3}
\end{figure}
\end{center}

In figure~\ref{fig:3}, we give examples of evolution for some of the relevant background quantities, i.e., for the slow-roll coefficient $\epsilon_H=-\dot H/H^2$ (panel a), inflaton amplitude $\phi$ (panel b), $T/H$ (panel c), $T$ (panel d), $Q$ (panel e) and energy densities (panel f), all as a function of the number of e-folds $N_e$,
for the benchmark parameters shown in table~\ref{tab1} and for the WI model given in table~\ref{tab2}. All evolutions are shown starting at $N_e=N_\star$ until the end of inflation, $N_e=N_{\rm end}$, where
for this example we have $N_{\rm end} - N_\star \simeq 46.2$ e-folds of inflation. 
Note that WI ends soon after the radiation energy density becomes dominant, as can be seen from figure~\ref{fig:3}(f).
This can also be easily seen when working with the slow-roll approximations for the background eqs.~(\ref{eq1})-(\ref{eq3}), from which we obtain
\begin{eqnarray}
\frac{\rho_r}{\rho_\phi} &=& \frac{\frac{\dot \phi^2}{2} + V}{\rho_r} \simeq \frac{ \frac{Q}{1+Q} \frac{\epsilon_H }{2}}{ 1+ \frac{\epsilon_H}{3(1+Q)}},
\label{rhoratio}
\end{eqnarray}
and where we also used eq.~(\ref{eq4}).
During WI we have the hierarchy $\dot \phi^2/2 \ll \rho_r \ll V$. With $Q$ increasing during WI, see fig.~\ref{fig:3}(e),
when $Q \gg 1$ towards the end of inflation,  then from eq.~(\ref{rhoratio}) we have that $\rho_r/\rho_\phi \sim 1/2$ when $\epsilon_H=1$.

The relevant cosmological observable quantities $n_s$ and $r$ are shown in figures~\ref{fig:4} and \ref{fig:5} as a function of the dissipation coefficient $Q$
evaluated at $N_\star$.  Here again we show only the case of thermalized perturbations, i.e., when considering
$n_\star=n_{BE}$ in the expression for the power spectrum 
in WI, eq.~(\ref{power_spectra}), since the results with $n_\star=0$ are indistinguishable from those with $n_\star=n_{BE}$ for the large $Q$ values considered here for this model, with a distintion between nonthermalized and thermalized inflaton perturbations only becoming more relevant in the $Q\lesssim 1$ region\footnote{Note that to determine which case, thermalization or not, for WI, requires a detailed study of nonequilibrium dynamics of the inflaton field in each model. Although such detailed analysis is outside the scope of the present work, recent studies indicate that thermalization of the inflaton perturbations tends to be favored
in the strong dissipation regime~\cite{ORamos:2025uqs,Berera:2025vsu}. }.

In figure~\ref{fig:4}(a) we show that for $100\lesssim Q_{\star}\lesssim 200$, the model prediction for the spectral tilt falls within the window of observation for both Planck/BICEP~\cite{BICEP:2021xfz} and ACT data \cite{ACT:2025tim,ACT:2025fju},
shown, respectively, by the cyan and orange shaded regions (for the one- and two sigma confidence levels (CL)). The tensor-to-scalar ratio,
shown in figure~\ref{fig:4}(b), becomes increasingly suppressed with larger $Q$, as expected in general in WI. This is a consequence of the fact that dissipation and thermal effects tend to enhance the scalar of curvature power spectrum relative to the tensor spectrum. 

\begin{figure}[!htb]
    \centering
    \includegraphics[width=0.5\linewidth]{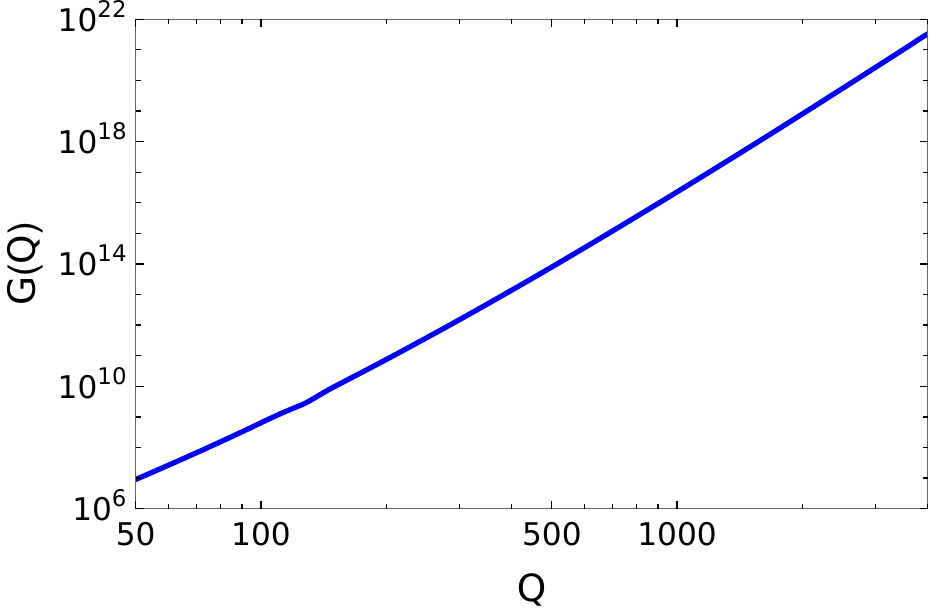}
    \caption{The numerical result for the function $G(Q)$ appearing in the WI power spectrum eq.~(\ref{power_spectra}) for the case of the radial direction brane inflation model and with dissipation coefficient $\Upsilon \propto T^3/\phi^2$.}
    \label{fig:6}
\end{figure}

The results for the different values of $Q$ when considered in the plane of $n_s$ and $r$ are shown in figure~\ref{fig:5}.
As described above, the model falls into the observation window for a wide range of $Q$ values in the large dissipation region.

{}Finally, for completeness, we show in figure~\ref{fig:6}  the comparison between the form of the $G(Q)$ function appearing in the power-spectrum~\eqref{power_spectra} for the model of radial brane WI with the dissipation coefficient $\Upsilon \propto T^3/\phi^2$ and within the range of $Q$ values of relevance here. $G(Q)$ was obtained here using the  numerical code \texttt{WI2easy}~\cite{Rodrigues:2025neh}.

\subsection{Angular brane WI}

Let us now study the case of angular brane WI.
The potential~\eqref{pot-angular} describes the angular motion of a $D3$-brane, where the brane motion is confined within the tip of a WDC --- a $S^3$. The scalar potential is sourced via F-terms, which contains the presence of a supersymmetric embedding---Kuperstein---of a stack of $D7$-branes. The potential is written as a function of one of the angular coordinates of a $S^3$ sphere. The other $5$ directions ($4$ angular and one radial) are stabilized by CY bulk corrections and assumed to be at their minimum position---an assumption generic to the study of brane inflation. 

The potential~\eqref{pot-angular} contains terms similar to natural inflation, \ie,  terms proportional to $\Lambda \left[1+\alpha_1\cos(\varphi/d)\right]$ plus additional terms. The appearance of extra sine and cosine terms helps to flatten the hilltop---forming a plateau.
The choice of parameters that we consider here and their origin in string theory is as follows. {}First, let us use \eqref{Angular_brane_pot1} to find the desired relation:
\begin{align}
    & \Lambda_1\propto \frac{2\kappa_4^2|A_0|a e^{-a\sigma}}{U^2},\quad \alpha_1\propto \frac{\varepsilon}{n\mu},\quad \alpha_2\propto \frac{\varepsilon^2}{n\mu^2}, \quad \Lambda_2\propto \frac{e^{-a\sigma}\varepsilon^{2/3}}{cn^2\gamma \mu^2},\quad \beta\propto \frac{e^{-a\sigma}\varepsilon^{5/3}}{cn^3\gamma \mu^3},
\end{align}
where $\gamma =\sigma T_{D3}/(3M_{\rm Pl}^2),\,U=2\sigma -\gamma \kappa_0$, with $\kappa_0$ the value of the K\"ahler potential at the tip of a DC and $\varepsilon$ is the radius of the $S^3$ sphere at the tip and $\mu^{2/3}$ denotes the distance between the $D3$-brane of the tip and the $D7$-brane (see figure~\ref{fig:illustration1} for a more geometric understanding). Similarly to the radial brane case,
considering $\sigma\sim \mathcal{O}(100),\,T_{D3}>1,\mu>1,\varepsilon<1$ and $an=2\pi$, we construct the following relations between the parameters: 
\begin{equation}
    \Lambda_2= \Lambda_1/2,\quad \alpha_2=\alpha_1/2,\quad \beta=\frac{\Lambda_1\alpha_1}{4},
\end{equation}
which gives us a simplified potential that is given by
\begin{equation}\label{ang_pot_simpl}
    V(\varphi)=\Lambda_1\left(1+\alpha_1\cos\frac{\varphi}{d}+\frac{\alpha_1}{2}\cos^2\frac{\varphi}{d}\right)+\frac{\Lambda_1}{2}\left(1-\frac{\alpha_1}{2}\cos\frac{\varphi}{d}\right)\sin^2\frac{\varphi}{d}.
\end{equation}
The (axion) effective decay constant retains its previous definition as $d=\varepsilon^{2/3}\sqrt{T_{D3} c}$. 

\begin{figure}[!htb]
    \centering
    \includegraphics[width=0.8\linewidth]{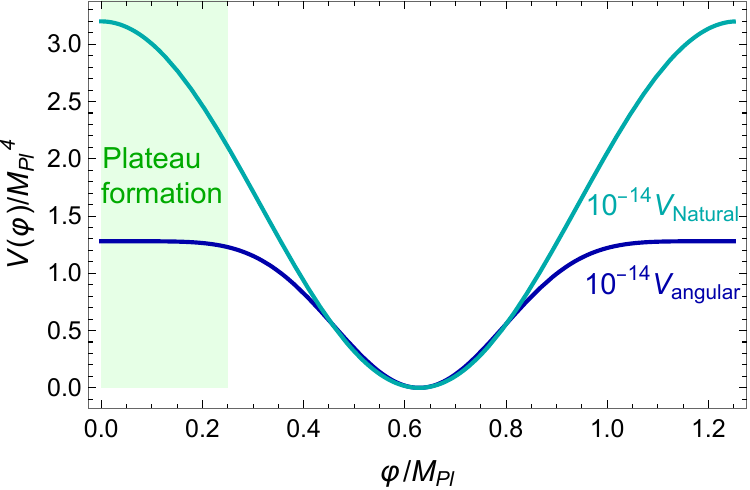}
    \caption{The angular-brane potential \eqref{ang_pot_simpl} when contrasted with the standard axion-like potential,  $V_{\text{Natural}}\propto \left(1+\cos\frac{\varphi}{d}\right)$. The parameters chosen for this plot are $d=0.2 M_{\rm Pl},\alpha_1=0.4$.}
    \label{fig:7}
\end{figure}

An illustration of the form of the potential~(\ref{ang_pot_simpl})
is shown in figure~\ref{fig:7}, where we also contrast its form with the standard potential of natural inflation for the same value for the effective decay constant $d$.

The relevant $60$ e-folds for CI occurs close to the branch of the natural inflation where the two potentials match. On the other hand, in WI, inflaton starts on the right-hand-side of the CI starting point for the same e-folds and ends near the minimum, displaying a shorter field displacement $\Delta \varphi$ (see  table~\ref{tab3}). For our choice of benchmark parameters shown in table~\ref{tab3}, the angular brane potential (\ref{ang_pot_simpl}) can sustain a successful period of both CI and WI.

\begin{table}[!htb]
\caption { Model parameters for the potential \eqref{ang_pot_simpl} that are used in the cases of CI and WI. }
\begin{center}
\centering
    \resizebox{0.8\textwidth}{!}{ 
    \begin{tabular}{| l | c | c | c | c | c | c | c | c |}
\hline
\cellcolor[gray]{0.9}   &\cellcolor[gray]{0.9} $\Lambda_1/M_{\rm Pl}^4$ &  \cellcolor[gray]{0.9} $\alpha_1$ &  \cellcolor[gray]{0.9} $d/M_{\rm Pl}$ & \cellcolor[gray]{0.9} $n_s$ & \cellcolor[gray]{0.9} $r$ & \cellcolor[gray]{0.9} $|\Delta\phi|/M_{\rm Pl}$ & \cellcolor[gray]{0.9} $Q_{\star}$ & \cellcolor[gray]{0.9} $C_T$ \\
\hline \hline
CI & $1\times 10^{-15}$ & $0.4$ & 0.1 & $0.90$ & $4.06\times 10^{-8}$ & $0.19$& $-$& $-$\\
\hline
 WI & $9.23\times 10^{-16}$  & $0.4$ & $0.1$ & $0.968$ & $2.37\times 10^{-8}$ & $0.20$ & $0.02$ & $7.06 \times 10^{-3}$\\
 \hline
 CI & $1.6\times 10^{-14}$  & $0.4$ & $0.2$ & $0.905$ & $3.32\times 10^{-7}$ & $0.38$ & $-$ & $-$\\
\hline
WI   & $6.42\times 10^{-15}$ & $0.4$ & $0.2$ & $0.969$ & $1.65\times 10^{-7}$ & $0.41$ & $0.02$ & $6.81 \times 10^{-3}$\\ 
\hline
CI & $5.88\times 10^{-14}$ & $0.4$ & $0.4$ & $0.904$ & $2.77\times 10^{-6}$& $0.77$& $-$ & $-$\\
\hline
 WI & $4.62\times 10^{-14}$  & $0.4$ & $0.4$ & $0.969$ & $1.19\times 10^{-6}$ & $0.84$ & $0.02$  & $6.59 \times 10^{-3}$\\
 \hline
\end{tabular}}
\end{center} 
\label{tab3}
\end{table}

In table~\ref{tab3} we also display our main results obtained for the case of the angular brane potential. {}For WI, the results presented refer to the case of dissipation that is linearly dependent on the temperature, $\Upsilon= C_T T$. 
Note that the dissipation constant $C_T$ can be related to the
parameters for our model with
interactions eq.~(\ref{Lintaxion}),  giving for us here (by comparing, e.g., with the similar result from ref.~\cite{Berghaus:2024zfg})
\begin{equation}
C_T \sim 8 \pi^3 N (N-1)^2g_{YM}^2 \frac{m^2}{d^2},
\label{CT}
\end{equation}
where we have assumed a general gauge group $SU(N)$
and the presence of light fermions with mass $m \ll T, d$.

The dynamics and perturbations in WI presented here are also evaluated assuming $n_\star=0$ in the power spectrum for WI, eq.~(\ref{power_spectra}), which has been shown to be appropriate when $Q\lesssim 0.1$, when working, e.g., with an axion type model with SM-like interactions~\cite{ORamos:2025uqs}.
In all the benchmark cases considered here, we have chosen parameters to allow to have sub-Planckian values for both $d$ and also for the field excursion. The requirement of trustworthy supergravity description of the warped deformed conifold in turn enforces a sub-Planckian field excursion of brane-inflation as discussed in \cite{Baumann:2006cd}. It should be noted that it is always challenging for natural inflation type potentials, \ie, potentials with a cosine dominated term,
in CI to match data unless super-Planckian axion decay constants are considered \footnote{Note that super-Planckian axion decay constants are generally difficult to realize in controlled string theory compactifications, often conflicting with the Weak Gravity Conjecture (WGC) \cite{Arkani-Hamed:2006emk}.}. All the CI cases with sub-Planckian axion decay constants presented in table~\ref{tab3} produces a too red-tilted spectral index and a higher value of the tensor-to-scalar ratio compared to what observation allows. However, the examples tabulated in table~\ref{tab3} show that WI can be fully consistent with the observations and also allowing for both an effective  decay constant and inflaton field excursion that are sub-Planckian. 

\begin{center}
\begin{figure}[!htb]
\subfigure[]{\includegraphics[width=4.7cm]{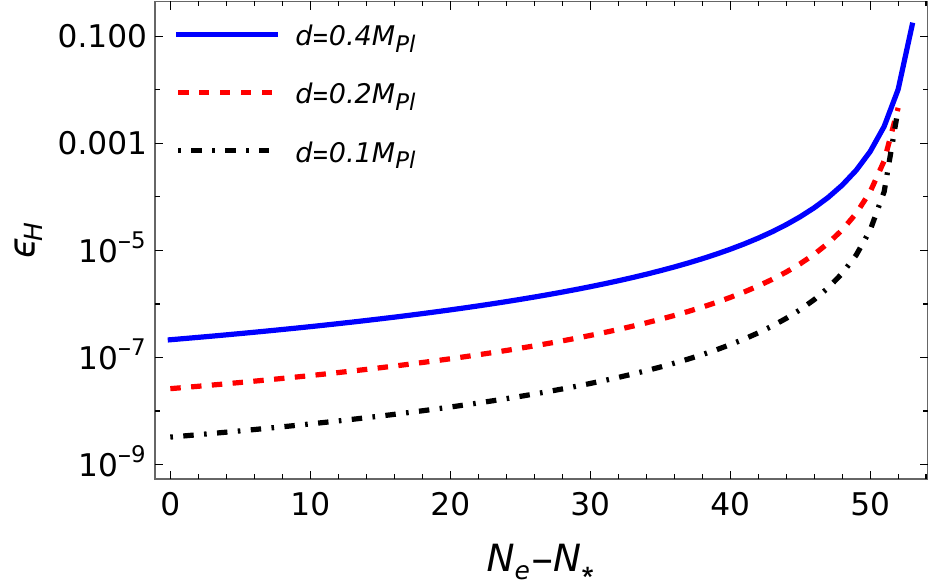}}
\subfigure[]{\includegraphics[width=4.6cm]{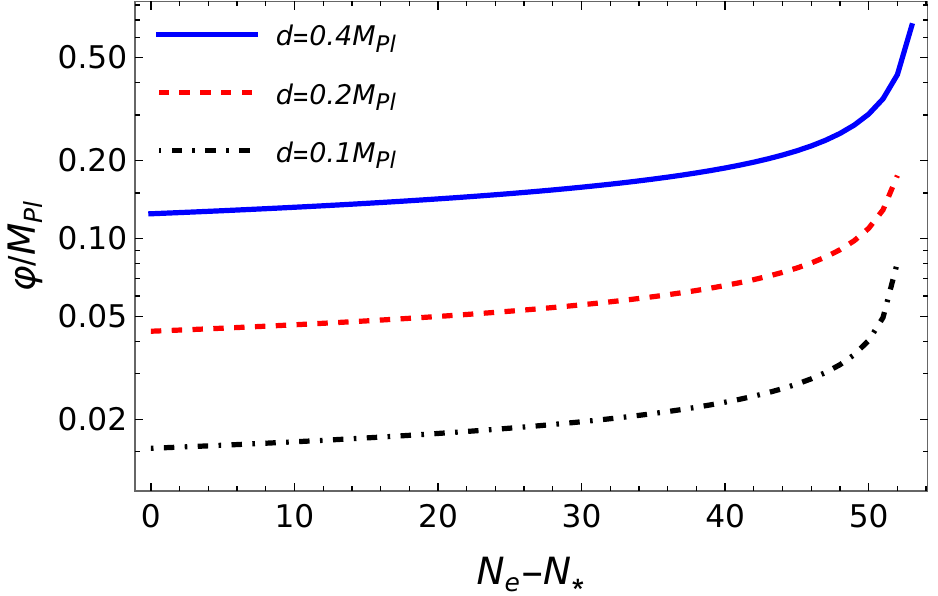}}
\subfigure[]{\includegraphics[width=4.8cm]{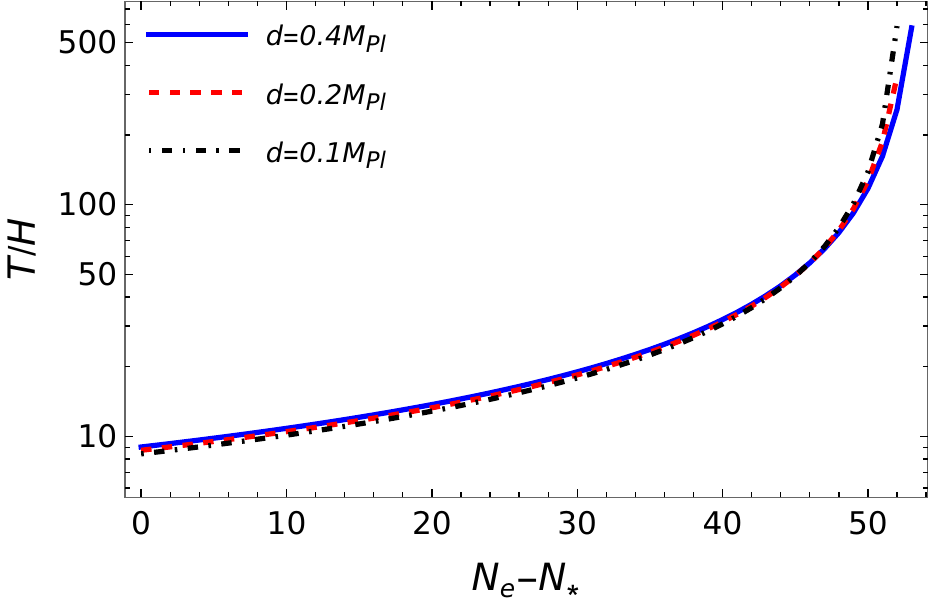}}
\subfigure[]{\includegraphics[width=4.8cm]{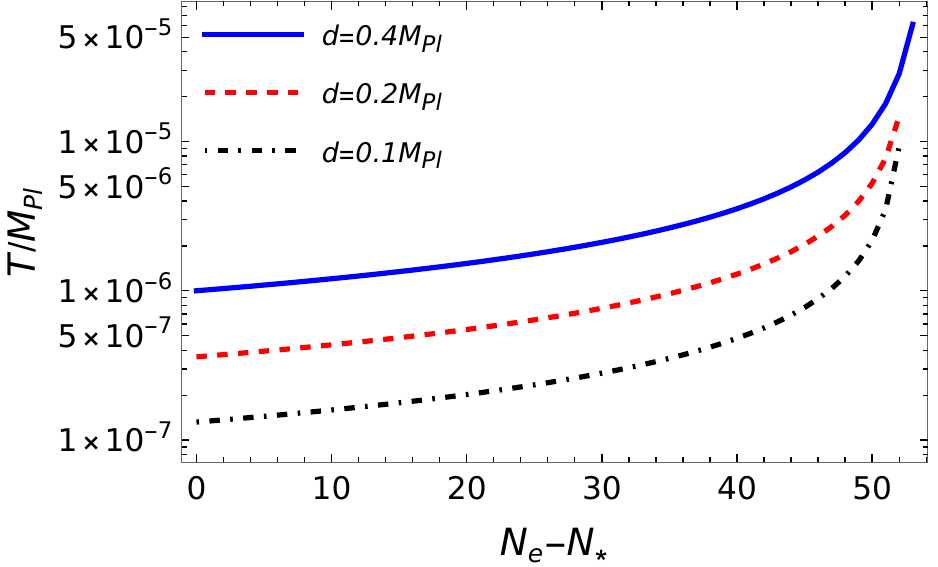}}
\subfigure[]{\includegraphics[width=4.6cm]{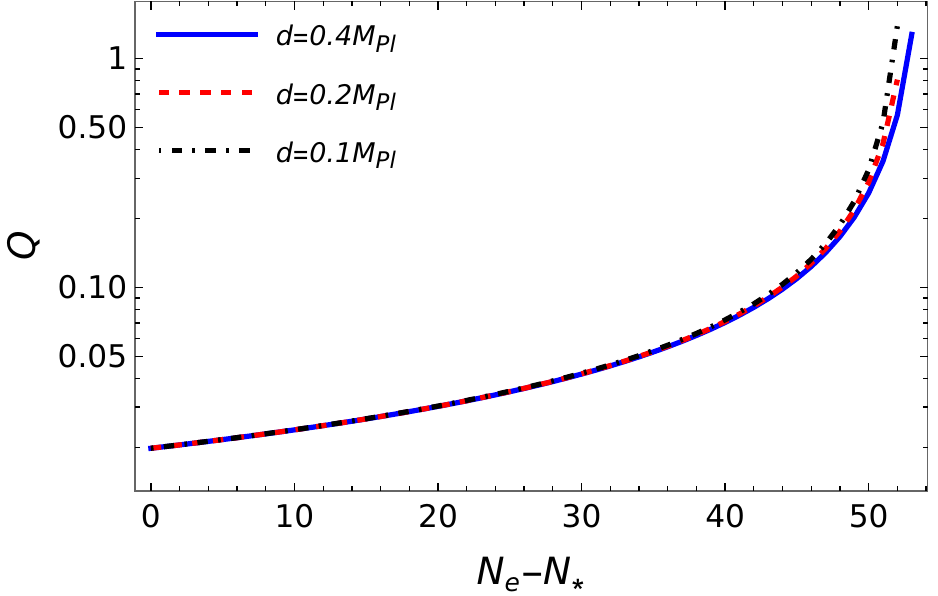}}
\hspace{0.5cm}
\subfigure[]{\includegraphics[width=4.6cm]{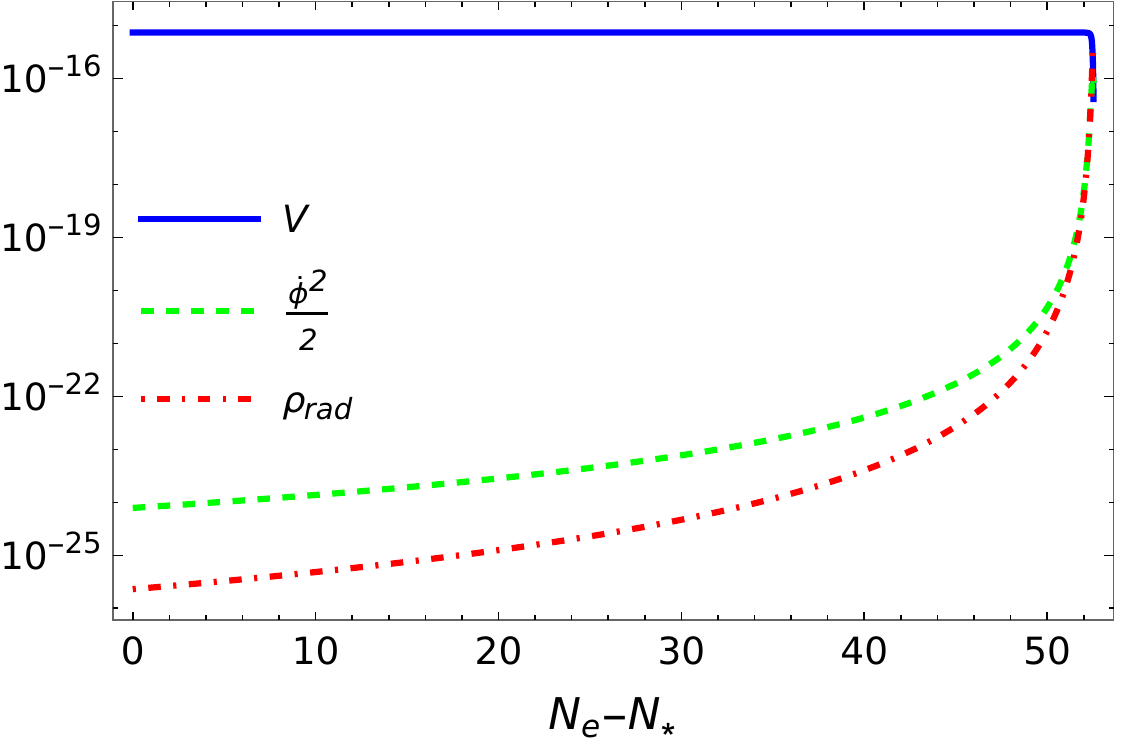}}
\caption{Evolution for the slow-roll coefficient $\epsilon_H$ (panel a), inflaton amplitude $\varphi$ (panel b), $T/H$ (panel c), $T$ (panel d) and $Q$ (panel e), for the three angular brane WI models given in table~\ref{tab3}. The energy densities (in units of the reduced Planck mass
$M_{\rm Pl}$) and shown in panel f, refers to the case of $d=0.1M_{\rm Pl}$.}    
\label{fig:8}
\end{figure}
\end{center}

Like in the previous example of radial brane WI, in figure~\ref{fig:8} we give the evolution for some of the relevant background quantities, i.e., for the slow-roll coefficient $\epsilon_H=-\dot H/H^2$ (panel a), inflaton amplitude $\varphi$ (panel b), $T/H$ (panel c), $T$ (panel d), $Q$ (panel e), all as a function of the number of e-folds $N_e$,
for the benchmark parameters shown in table~\ref{tab3} and the three values of $d$ chosen there. {}For the energy densities (panel f) we have chosen to show them for the case of $d=0.1 M_{\rm Pl}$ to not clutter the plot and the other cases are very similar. All evolutions are shown starting at $N_e=N_\star$ until the end of inflation, $N_e=N_{\rm end}$, where
here we have that $N_{\rm end} - N_\star \simeq 53.4$ e-folds (for $d=0.4 M_{\rm Pl}$), $52.9$ e-folds (for $d=0.2 M_{\rm Pl}$), and  $52.5$ e-folds (for $d=0.1 M_{\rm Pl}$). 
As in the case of the radial brane WI, see eq.~(\ref{rhoratio}), inflation here ends soon after the radiation energy density overtakes the potential energy density for the inflaton field.

\begin{center}
\begin{figure}[!bth]
\subfigure[]{\includegraphics[width=7.5cm]{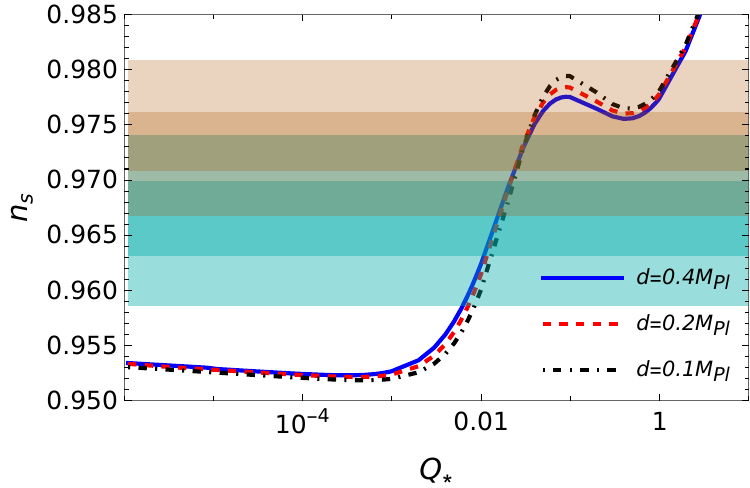}}
\subfigure[]{\includegraphics[width=7.5cm]{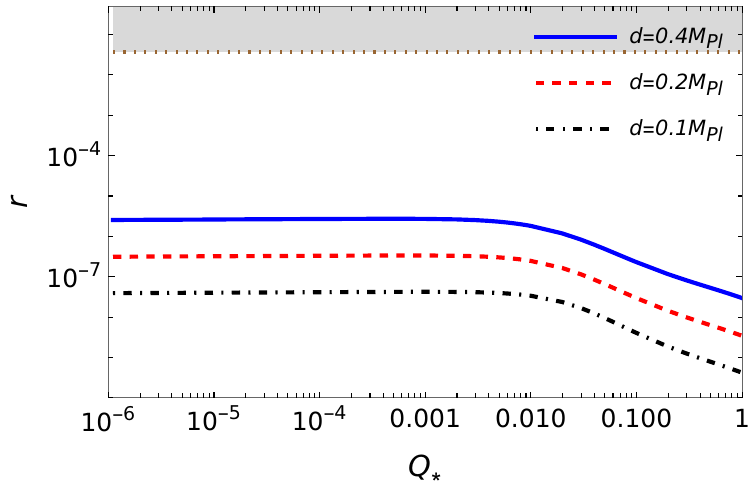}}
\caption{Results for $n_s$ (panel a) and $r$ (panel b) as  a function of $Q_{\star}$ (at the Hubble radius crossing point) for the case of the angular brane inflation with a dissipation coefficient linear in the temperature. The gray shaded region above the horizontal dotted line shown in panel (b) indicates the region above the upper bound value for the tensor-to-scalar ratio~\cite{BICEP:2021xfz}, $r< 0.036$.}    
\label{fig:9}
\end{figure}
\end{center}


\begin{figure}[!htb]
    \centering
    \includegraphics[width=0.5\linewidth]{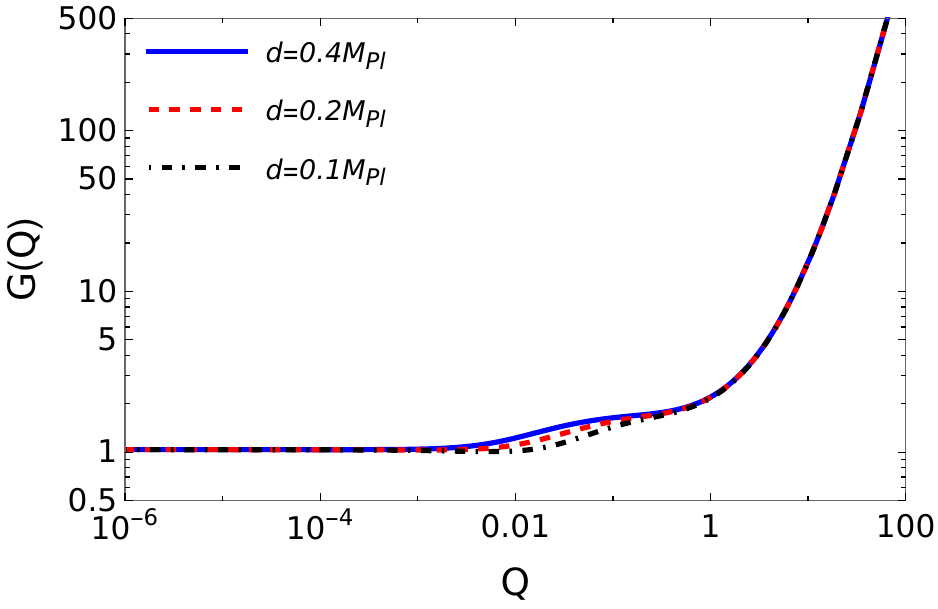}
    \caption{The function $G(Q)$ appearing in the WI power spectrum eq.~(\ref{power_spectra}) for the case of the angular brane inflation model with a dissipation coefficient $\Upsilon \propto T$.}
    \label{fig:10}
\end{figure}

\begin{figure}[!htb]
    \centering
    \includegraphics[width=0.63\linewidth]{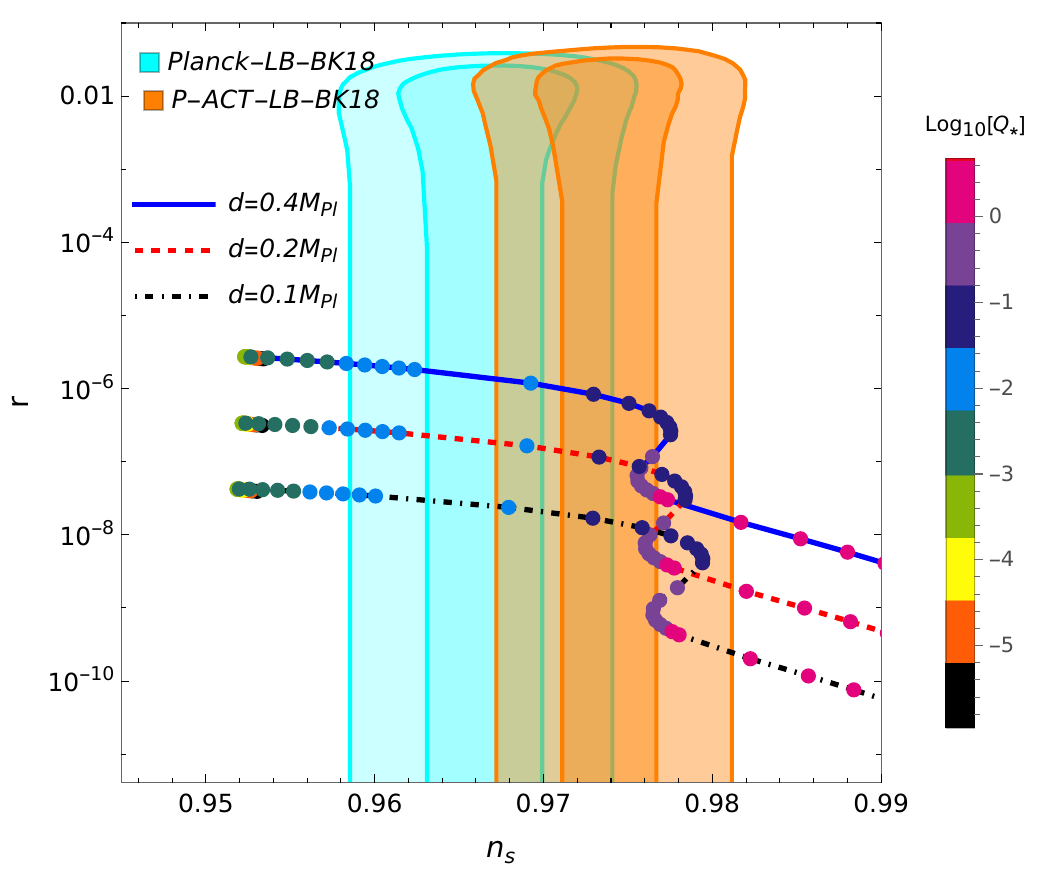}
    \caption{Results for the angular brane model in WI, with a dissipation coefficient that is linear in the temperature, shown in the plane $n_s-r$ against Planck and ACT data.}
    \label{fig:11}
\end{figure}

Analogously to what we have shown for the case of radial brane inflation, in figures~\ref{fig:9}-\ref{fig:11} we show
the relevant cosmological perturbation quantities  as a function of the dissipation coefficient $Q$
evaluated at $N_\star$,  obtained for the three different values of $d$ shown in table~\ref{tab3} and taking $n_\star=0$ in eq.~(\ref{power_spectra}).
In figure~\ref{fig:9} we show the results for the spectral tilt (panel a) and tensor-to-scalar ratio (panel b). These results show that the angular brane WI can accommodate a large range of $Q_{\star}$ for which the model falls in the allowed observation window of both Planck and ACT data. Once again, the presence of dissipation suppresses the value of the tensor-to-scalar ratio.
In figure~\ref{fig:10} we show the function $G(Q)$ for the angular brane WI models of table~\ref{tab3} and in figure~\ref{fig:11} we show the results for $n_s$ and $r$ 
in terms of the one- and two-sigma CL contours from both Planck/BICEP and ACT. {}For the model parameters considered, WI here is favored in the weak dissipation regime (where $Q \ll 1$) for the Planck data and being able reach up to $Q \sim 1$ for the combination of Planck and ACT data.

Before closing this section, we comment on the expected level of primordial non-Gaussianity in the models analyzed above. Using the formalism and numerical methods developed in Ref.~\cite{Bastero-Gil:2014raa}, we estimate, for the benchmark parameters listed in table~\ref{tab2}, that the radial brane inflation model gives $f_{\rm NL}^{\rm warm} \simeq 5.5$. {}For the angular brane inflation benchmarks listed in table~\ref{tab3}, we obtain $f_{\rm NL}^{\rm warm}\simeq 3.7,\quad 4.0,\quad\text{and}\quad 4.4$ for $d/M_{\rm Pl}=0.1, \;0.2$, and $0.4$, respectively. These estimates indicate that all the benchmark models predict non-Gaussianities of order unity, without any parametrically large enhancement.
{}For comparison, the dedicated Planck analysis employing the warm-inflation bispectrum template found $f_{\rm NL}^{\rm warm}=-48\pm 27$ at $68\%$ confidence level~\cite{Planck:2019kim}. Our estimates are therefore compatible with the absence of a statistically significant primordial non-Gaussian signal. We caution, however, that the WI template employed in the Planck analysis was derived for the strong dissipative regime and is associated with theoretical estimates of $| f_{\rm NL}^{\rm warm}|$ of order ten~\cite{Moss:2007cv}, whereas the angular brane inflation benchmarks considered here evolve in the weak dissipative regime. Since both the amplitude and the shape of the WI bispectrum can differ between the strong and weak dissipative regimes, a dedicated analysis using an appropriate weak-dissipation template would be required for a direct comparison with the CMB data.

\section{Discussion and conclusion}
\label{sec:5}

In this work, we investigated warm inflation (WI) within the context of brane-motivated models. 
Our approach differs fundamentally from previous warm brane inflation 
studies~\cite{Bastero-Gil:2011zxb,Cid:2007fk,delCampo:2007cia,Setare:2014qea,Kamali:2019xnt}. 
Notably, our setup does not rely on an anti-brane; instead of Coulomb interactions dominating 
the scalar potential, the primary contributions are induced via moduli stabilization. {}Furthermore, 
we provide a distinct analysis of both radial and angular brane inflation within a 
warped deformed conifold (WDC). For radial brane inflation, the inflaton is identified with 
the radial distance between the UV end of the warped throat (housing the $D7$-brane) and 
the IR end. For angular brane inflation, the brane sits at the tip of the throat, and the 
inflaton corresponds to one of the WDC's angular coordinates. Both the inflationary models have sub-Plackian field excursions and sub-Plankican decay constant in case of angular brane inflation. Through sub-Planckian field excursion, the models can also safely bypass the controversy of swampland program especially the swampland distance conjecture of \cite{Ooguri:2006in}. Crucially, while cold 
inflation (CI) fails to meet Planck and ACT observational constraints for this parameter space, 
we demonstrate that WI successfully brings both models into a phenomenologically viable regime. 
Additionally, all model parameters are fully motivated by their origins in string theory.

We examined two distinct dissipation mechanisms tailored to these models. {}For radial brane 
inflation, we assume that the inflaton couples to heavy intermediate particles that subsequently decay 
to lighter particles, producing a dissipation rate proportional to $T^3/\phi^2$. {}For angular brane 
inflation, the inflaton descends from a complex scalar field: its phase acts as the inflaton (axion), 
while the real part corresponds to the conifold modulus. We constructed an interaction Lagrangian 
where the conifold modulus is coupled to the gauge kinetic term ($FF$), and the axionic component 
couples to the topological term ($F\tilde{F}$). This Yang-Mills coupling induces a 
dissipation rate linear in temperature, driven by sphaleron processes in a thermal bath 
with non-chiral fermions---a mechanism well-motivated by type-IIB superstring theory.

{}Finally, we have presented that the radial brane inflation facilitated by the two-stage dissipation mechanism manifest itself as the first realization in the literature to have accessed a very strong dissipation regime of $100\lesssim Q\lesssim 160$, 
at the same time remaining observationally viable. {}For angular brane 
inflation, the linear dissipation mechanism allows the realization of a WI framework with a 
sub-Planckian effective decay constant, rendering this scenario highly attractive for rigorous 
string-theoretic model building.

\begin{appendix}
    \section{Warped Deformed Conifold}\label{appA}

    Let us now describe the singular and deformed conifold (DC) in order to familiarize the reader about the formulas used in this article. A singular conifold is a cone like local structure of a non-compact CY threefold. It is defined as a hypersurface in $\mathbb{C}^4$. There are two sets of complex projective coordinates which help us to define a conifold:
 one is denoted by $\omega_i$ with $i=1,2,3,4$ and the second set is denoted by $z_A$. For both $\{\omega_i,z_i\}$ the defining equations are:
 \begin{align}
     \omega_1\omega_2-\omega_3\omega_4&=0,\nonumber\\
     \sum_{i=1}^4\left(z_i\right)^2=0.
 \end{align}
 Besides, $\{\omega_i,z_i\}$ are related via:
 \begin{align*}
     z_1&=\frac{1}{2}\left(\omega_1+\omega_2\right),\qquad z_2=\frac{1}{2i}\left(\omega_1-\omega_2\right),\\
     z_3&=\frac{1}{2}\left(\omega_3-\omega_4\right),\qquad z_4=\frac{1}{2i}\left(\omega_3+\omega_4\right).
 \end{align*}
 The metric of a conifold can be written in terms of six real coordinates, which include a radial direction $r$ and $5$ angular directions $\phi_1,\phi_2,\theta_1,\theta_2,\psi$. The base of the cone is $T^{1,1}$, which is topologically equivalent to $S^3\times S^2$. A DC is defined in $\mathbb{C}^4$ as:
\begin{equation}\label{conifold_constraints}
    \sum_{i=1}^{4}(z_i)^2=\varepsilon^2,
\end{equation}
where $\varepsilon$ is related to the radius of the $S^3$ at the tip of the DC. The complex coordinates $(\omega_A)$ are expressed in terms of the real coordinates as:
\begin{align}
    \omega_1&=r^{3/2}e^{\frac{i}{2}(\psi-\phi_1-\phi_2)}\sin\frac{\theta_1}{2}\sin \frac{\theta_2}{2},\\
    \omega_2&=r^{3/2}e^{\frac{i}{2}(\psi+\phi_1+\phi_2)}\cos\frac{\theta_1}{2}\cos \frac{\theta_2}{2},\\
    \omega_3&=r^{3/2}e^{\frac{i}{2}(\psi+\phi_1-\phi_2)}\cos\frac{\theta_1}{2}\sin \frac{\theta_2}{2},\\
    \omega_4&=r^{3/2}e^{\frac{i}{2}(\psi-\phi_1+\phi_2)}\sin\frac{\theta_1}{2}\cos \frac{\theta_2}{2}.
\end{align}

Two explicit supersymmetric $D7$-brane embeddings have been discovered by Ouyang \cite{Ouyang:2003df} and Kuperstein \cite{Kuperstein:2004hy}. In case of Ouyang embedding, the functional $f(z|\omega)$ takes the form of:
\begin{equation}\label{Ouyang_embedding}
    f(\omega_i)=1-\frac{\omega_1}{\mu},
\end{equation}
where $\mu^{2/3}$ denotes the distance between $D3$-brane at the tip and the $D7$-brane at the UV range of the conifold. In the derivation of the  K\"ahler potential, the open string moduli dependent part of \eqref{kahler_pot_gen} is expressed as:
\begin{equation}
    k(\omega_i,\bar{\omega}_i)=r^2=\left(\sum_{i=1}^4|\omega_i|^2\right)^{2/3}
\end{equation}

On the other hand, Kuperstein embedding is denoted via:
\begin{equation}\label{Kuperstein_embedding}
    f(z_i)=1-\frac{z_1}{\mu}.
\end{equation}
Close to the tip, the K\"ahler potential of the DC looks like \cite{Kaluza:1921tu,Gukov:1999ya}:
\begin{equation}\label{Kahler_DC}
k(z,\bar{z})=k_0+c\,\varepsilon^{-2/3}\left(\sum_{i=1}^{4}|z_i|^2-\varepsilon^2\right),
\end{equation}
where $c=\frac{2^{1/6}}{3^{1/3}}\simeq 0.77$. $c$ has dimension of $\text{Mass}^2$. We will choose the set of coordinates $(z_1,z_2,z_3)$ as the independent one and impose $z_4^2=\varepsilon^2-\sum_{i=1}^3z_i^2$ from \eqref{conifold_constraints} and following~\cite{Pajer:2008uy}, we also assume $z_i=\bar{z}_i=x_i$.
\end{appendix}

\section*{Acknowledgments}

D.C. thanks Ivonne Zavala for useful discussions and comments. 
R.O.R. acknowledges financial support by research grants from Conselho
Nacional de Desenvolvimento Cient\'{\i}fico e Tecnol\'ogico (CNPq),
Grant No. 307286/2021-5, and from Funda\c{c}\~ao Carlos Chagas Filho
de Amparo \`a Pesquisa do Estado do Rio de Janeiro (FAPERJ), Grant
No. E-26/200.415/2026.


\providecommand{\href}[2]{#2}\begingroup\raggedright\endgroup



\end{document}